\documentclass[journal,twocolumn,10pt]{IEEEtran}
\usepackage{graphicx}
\usepackage{verbatim}
\usepackage{upgreek}
\usepackage{amssymb,amsmath}
\usepackage{color}
\usepackage{epstopdf}
\usepackage{bm}
\usepackage{cite}
\usepackage{array,color}
\usepackage{algorithm}
\usepackage{algpseudocode}
\usepackage{amsmath}
\usepackage{amsfonts}
\usepackage{amsthm}
\usepackage{graphics}
\usepackage{epsfig}
\usepackage{soul}
\usepackage{booktabs}
\usepackage{multirow}
\usepackage{stfloats}
\soulregister\cite7
\soulregister\ref7
\usepackage{subfigure}
\usepackage{tabularx}
\usepackage{makecell}
\usepackage{stfloats}
\usepackage{graphicx}
\usepackage{gensymb}
\usepackage{caption}
\renewcommand\appendix{\setcounter{secnumdepth}{3}}

\usepackage{etoolbox}
\newcommand{\tabincell}[2]{\begin{tabular}{@{}#1@{}}#2\end{tabular}} 
\makeatletter
\patchcmd{\@makecaption}
{\scshape}
{}
{}
{}
\makeatletter
\patchcmd{\@makecaption}
{\\}
{.\ }
{}
{}
\makeatother

\ifCLASSINFOpdf
\else
\fi
\allowdisplaybreaks
\begin{document}

\title{ \huge Integrated Sensing, Communication, and Computing in Multi-Tier Systems: Joint Hybrid Beamforming Design and Computation Resource Allocation}

\author{Peng Liu, Zesong Fei,~\IEEEmembership{Senior~Member,~IEEE}, Xinyi Wang,~\IEEEmembership{Member,~IEEE},  Qiao Qi,~\IEEEmembership{Member,~IEEE},\\ Zhaohui Yang,~\IEEEmembership{Member,~IEEE}, Meng Hua,~\IEEEmembership{Senior~Member,~IEEE}, and Arumugam Nallanathan,~\IEEEmembership{Fellow,~IEEE}

	\thanks{Peng Liu, Zesong Fei, and Xinyi Wang are with the School of Information and Electronics, Beijing Institute of Technology, Beijing 100081, China (e-mail: bit\_peng\_liu@163.com, feizesong@bit.edu.cn, bit\_wangxy@163.com).}
			\thanks{Qiao Qi is with the School of Information Science and Technology, Hangzhou Normal University, Hangzhou 311121, China (email: qiqiao@hznu.edu.cn). }
				\thanks{Zhaohui Yang is with the College of Information Science and Electronic Engineering, Zhejiang University, Hangzhou, 310027, China (email: yang\_ zhaohui@zju.edu.cn). }
				\thanks{Meng Hua is with the Department of Electrical and Electronic Engineering, Imperial College London, London SW7 2AZ, UK. (e-mail: m.hua@imperial.ac.uk).}
				\thanks{Arumugam Nallanathan  is with the School of Electronic Engineering and Computer Science, Queen Mary University of London, E1 4NS London, U.K., and also with the Department of Electronic Engineering, Kyung Hee University, Yongin-si, Gyeonggi-do 17104, South Korea (E-mail: a.nallanathan@qmul.ac.uk).}

}

\maketitle

\begin{abstract}
This paper proposes a novel integrated sensing, communication, and computing (ISCC) framework over a cloud–edge–device collaborative architecture, where passive sensing is enabled by reusing uplink offloading signals to extract sensing information directly at the edge without incurring additional transmission overhead. Nevertheless, such signal reuse introduces an inherent tradeoff between communication efficiency and sensing coverage. To address this challenge, we adopt a hybrid beamforming architecture under practical hardware constraints. In addition, the integration of sensing tasks creates significant resource contention at the mobile edge computing (MEC) server, where latency-sensitive device tasks and computation-intensive sensing inference tasks compete for limited processing capacity. To alleviate this computation burden, we introduce a split inference mechanism that strategically partitions intelligent sensing tasks between the edge and the cloud. Building upon this framework, we formulate a joint optimization problem to minimize the average computation latency of all device tasks subject to strict sensing performance constraints. To tackle the high non-convexity of the formulated problem, we develop an efficient alternating optimization algorithm. In particular, we design a two-layer framework to jointly determine the optimal DNN splitting point and computation resource allocation and employ a weighted minimum mean square error (WMMSE)-based approach with manifold optimization for hybrid beamforming design. Numerical results demonstrate that the proposed framework achieves a superior tradeoff between sensing accuracy and computation latency compared to the benchmark schemes.

\vspace{2ex}
\end{abstract}

\IEEEpeerreviewmaketitle
\section{Introduction}
With the evolution toward sixth-generation (6G) wireless networks, emerging applications such as extended reality (XR), intelligent transportation, smart homes, and immersive human–machine interaction are expected to be natively supported \cite{iot1}. These applications impose stringent latency and computational requirements, which pose significant challenges to the limited processing capability and energy budget of end devices. To address this limitation, mobile edge computing (MEC) has been widely recognized as a key enabling paradigm by offloading computation tasks to edge servers in close proximity to end devices, thereby reducing execution latency and device-side computation burden \cite{mecto}. Motivated by this paradigm, extensive research efforts have focused on MEC-enabled wireless networks \cite{mec2,mec3,mec4}. In \cite{mec2}, the authors investigated multi-user MEC systems enabled by multiple-input multiple-output (MIMO) techniques, where beamforming vectors and computation resources were jointly optimized to improve energy efficiency. To cope with the limited computation capacity of MEC servers, the authors in \cite{mec3} and \cite{mec4} investigated cloud–edge–end collaborative computing architectures, in which task partitioning and computation resource allocation were optimized to further reduce task processing latency.

Beyond the provision of computation capability, future wireless networks are also expected to possess environmental sensing functionality to support a wide range of intelligent applications, such as autonomous driving and XR. To this end, integrated sensing and communication (ISAC) has emerged as a promising paradigm by unifying sensing and communication through shared spectral and hardware resources, enabling efficient spectrum utilization and reduced hardware cost \cite{ISACLi}. However, the devices with limited computation capability are generally incapable of processing the massive volumes of sensing data collected under ISAC in real time, especially for intelligent sensing tasks, such as complex target detection and classification based on deep neural network (DNN) models \cite{Alexnet1}. To address this issue, integrating ISAC with MEC enables sensing data offloading to nearby MEC servers, thereby reducing device-side computation burden and supporting real-time intelligent sensing \cite{isccsurvey}.  Along this line, the authors in \cite{ISCC2} exploited ISAC signals to simultaneously perform environmental sensing and sensing data offloading. Through joint beamforming and computation resource allocation design, the proposed scheme reduced the processing latency and energy consumption of sensing tasks. In \cite{UAVXu}, an \textcolor{black}{autonomous aerial vehicle (AAV)}-enabled ISAC–MEC framework was considered, where sensing data collected by AAVs were transmitted to MEC servers for processing, and the AAV trajectory and beamforming were jointly optimized to maximize the throughput. To improve the reliability of task offloading, Huang \textit{et al.} proposed a short-packet sensing data offloading scheme and optimized the packet length to ensure a low decoding error probability while reducing device energy consumption \cite{Huang2024}. The author in \cite{lptcom2} investigated joint radio and computation resource allocation in ISCC systems and minimized sensing task completion latency under detection probability constraints.  In \cite{peng2024tcom,LP2,lptvt1}, this framework was further extended to a three-tier cloud–edge–end architecture, where ISAC beamforming were designed to enable hierarchical task processing and further reduce sensing task processing latency.  

{  
    \begin{table*}[!ht]
		\renewcommand{\arraystretch}{1.3}
		\caption{Comparison between our work and the existing ISCC works}
		\label{1}
		\centering
		\renewcommand{\arraystretch}{1.3}
	\color{black}\begin{tabular}{|c|c|c|c|c|c|c|c|c|}
			\hline
			\bfseries Reference  & \tabincell{c}{\bfseries ISAC} &  \tabincell{c}{\bfseries Cloud-edge- \\ \bfseries device}   & \tabincell{c}{\bfseries MIMO } & \tabincell{c}{\bfseries Hybrid beamforming }& \tabincell{c}{\bfseries DNN split inference} & \tabincell{c}{\bfseries Uplink sensing \\ \bfseries by offloading signal}& \tabincell{c}{\bfseries Cooperative sensing } \\
			\hline
			\cite{ISCC2} & \checkmark& &\checkmark & & & &  \\
			\hline
			\cite{UAVXu} &\checkmark& & & & & &  \\
			\hline
			\cite{Huang2024} &\checkmark& & \checkmark & & & &  \\
			\hline
				\cite{lptcom2} &\checkmark &&\checkmark & & & &  \\
			\hline
			\cite{peng2024tcom} &\checkmark&\checkmark& \checkmark & & & & \\
			\hline
				\cite{LP2} &\checkmark&\checkmark &\checkmark & &\checkmark & &  \\
				\hline
					\cite{lptvt1} &\checkmark&\checkmark &\checkmark & & & &  \\
					\hline
					
			Our work &\checkmark&\checkmark &\checkmark&\checkmark&\checkmark&\checkmark&\checkmark \\
			\hline
		\end{tabular}
		
\end{table*}}

Despite the significant potential of integrating MEC with ISAC, directly offloading sensing data inevitably occupies valuable uplink bandwidth and may compete with device computation data offloading for limited wireless resources. Fortunately, recent advances in communication-signal-based sensing \cite{random1,MECS} and bistatic sensing \cite{uplink1,ISCCPS} techniques have shown that reliable environmental sensing can be achieved by reusing the received communication data signals \cite{random3,MECS}. Inspired by these observations, \textcolor{black}{\textit{this paper considers a novel integrated sensing, communication, and computing (ISCC) architecture, in which the base station (BS) performs  passive sensing by exploiting the uplink offloading signals transmitted by multiple devices, such that sensing data are generated at the edge, thus avoiding the need for additional sensing data offloading.}} Nevertheless, this architecture also introduces new challenges. On the one hand, efficient sensing and computation offloading require sufficient spatial degrees of freedom at both devices and the BS to enable flexible control of beam directions and sensing resolution. However, in practical systems, large-scale antenna arrays are often constrained by a limited number of radio-frequency (RF) chains and hardware cost, rendering fully digital beamforming impractical. \textcolor{black}{On the other hand, although migrating sensing tasks from the devices to the edge can effectively alleviate device-side computation burden, it also requires MEC servers to simultaneously handle computation offloading and sensing-related processing, thereby increasing their computation load. In particular, DNN-based sensing inference introduces non-negligible computational workloads. For example, recent millimeter-wave radar perception networks require tens to hundreds of GFLOPs per inference \cite{dnnload}. Therefore, efficient allocation of MEC/cloud computation resources and appropriate DNN execution strategies are essential to balance the heterogeneous workloads in ISCC systems.}

To address these challenges, hybrid beamforming \cite{cheng2021} and split inference \cite{DNNsplit2} have been regarded as two promising enabling techniques for ISCC systems. 
Hybrid beamforming enables efficient spatial processing by jointly designing analog and digital beamformers under limited RF chains, providing a flexible tradeoff between sensing performance and communication efficiency \cite{wangtcom2022}. 
Meanwhile, split inference enables adaptive partitioning of DNN models across edge and cloud layers, allowing computation workloads to be distributed among heterogeneous computing tiers and reducing the computation burden at individual nodes \cite{dnnmec1}.
\textcolor{black}{Although our previous studies \cite{LP2} have demonstrated the effectiveness of split inference in three-tier ISCC architectures, they mainly focused on active sensing paradigms, where sensing tasks require additional communication resources, and adopted fully digital beamforming architectures \cite{lptcom2, peng2024tcom, LP2,lptvt1}. To overcome these limitations, this work investigates a passive uplink signal reuse sensing framework with hybrid beamforming, where uplink communication signals from multiple devices are exploited for collaborative sensing without additional sensing transmissions. Under this framework, sensing inference and device computation tasks share MEC and cloud computing resources, resulting in stronger coupling among communication, sensing, and computation processes. Therefore, the joint design of hybrid beamforming and multi-tier computation resource allocation under passive uplink sensing remains a challenging and insufficiently investigated problem in ISCC systems.}

Based on the above analysis, we incorporate \textcolor{black}{uplink sensing}, split  inference and hybrid beamforming into the cloud–edge–device collaborative computing system with the device computation tasks jointly executed by the devices, the MEC server, and the cloud server. In addition to exploiting the uplink offloading signals from multiple devices for cooperative target detection, the BS collaborates with the cloud server to accomplish DNN-based sensing inference tasks. \textcolor{black}{The differences between our work and existing ISCC works are summarized in Table I}, and the main contributions of this paper are summarized as follows:

\begin{itemize}
	\item \textcolor{black}{We propose a novel ISCC-enabled cloud--edge--device hierarchical computing architecture, in which uplink computation offloading signals are exploited to enable cooperative sensing without introducing dedicated sensing transmissions. Moreover, by incorporating cloud--edge collaborative DNN inference, the proposed framework effectively supports computation-intensive sensing tasks while alleviating the computation burden at the MEC server.}

	\item \textcolor{black}{We systematically investigate the intrinsic coupling among sensing, communication, and computation in the considered ISCC system. By characterizing the impact of hybrid beamforming, DNN split inference, and heterogeneous computation resources on both sensing performance and task execution latency, we formulate a joint optimization problem that minimizes average device task execution latency under explicit sensing performance constraints.}
	
	\item \textcolor{black}{To tackle the nonconvexity of the resulting joint optimization problem, we develop an efficient alternating optimization algorithm. Specifically, we propose a two-layer optimization framework to jointly determine the DNN splitting point and multi-tier computation resource allocation, where closed-form optimal solutions for computation resource allocation are derived via the Karush--Kuhn--Tucker (KKT) conditions, and the optimal DNN partitioning point is obtained through one-dimension search. Furthermore, we develop a weighted minimum mean square error (WMMSE)-based optimization approach combining  manifold optimization to jointly optimize the hybrid transmit beamformer and combiner. Finally, leveraging the equalization principle of
	task execution latency, we obtain the optimal closed-form partition
	ratio for device computation tasks.}
	
\end{itemize}

\textcolor{black}{Through numerical results, we demonstrate the performance advantages of the proposed scheme in reducing device task execution latency. Compared to benchmark schemes without split inference, the proposed scheme achieves lower computation latency and a better tradeoff between latency and detection probability. Furthermore, with additional sensing functionality and hybrid array architectures, the proposed scheme incurs only marginal latency degradation compared to three-tier computing systems without sensing requirements and fully digital array-based benchmarks.}

 The remainder of this paper is organized as follows. Section II presents the system model and the problem formulation. Section III proposes an alternating optimization algorithm  to solve the formulated problem. Simulation results are provided in Section IV to demonstrate the effectiveness of the proposed scheme. Finally, Section V concludes the paper.


\section{System Model}

\begin{figure}[!t]
	\centering
	\includegraphics[width=2.4in]{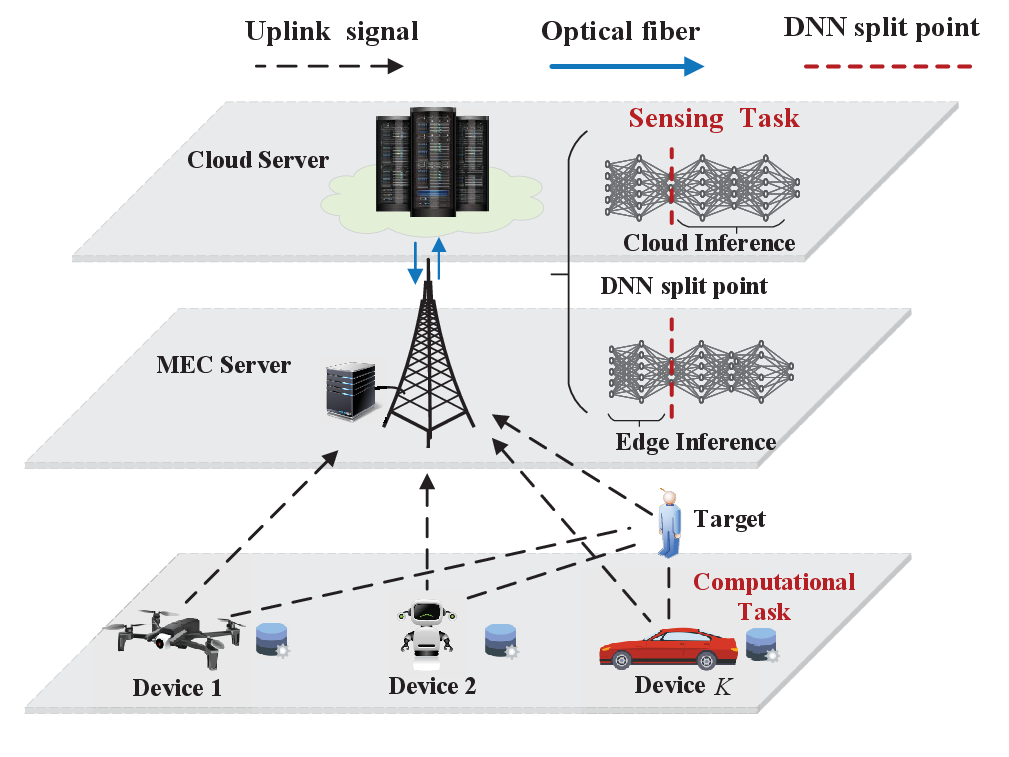}
	\caption{\textcolor{black}{The ISCC-enabled three-tier cloud-edge-device  network architecture.}}
	\label{fig:sys}
\end{figure}
As illustrated in Fig. 1, we consider a cloud–edge–device collaborative computing system consisting of one cloud server, one BS, and $K$ devices indexed by $\mathcal{K}=\{1,\ldots,K\}$. The BS is equipped with an MEC server and  connected to the cloud server via optical fiber. Each device has a computation-intensive task, which is divisible (e.g., image or video processing) \cite{mec3,mec4}, and can be partitioned into multiple segments for parallel processing. To fully exploit the computing capabilities at different tiers and reduce the overall execution latency, each device adopts a partial offloading strategy, splitting its task into three portions executed locally, at the MEC server, and at the cloud, respectively. We define $\alpha_k$, $\beta_k$, and $\gamma_k$ as the task partition ratios \textcolor{black}{for locally, the MEC server, and at the cloud, respectively,}  satisfying $\alpha_k+\beta_k+\gamma_k=1$. 

Both the BS and the devices employ a hybrid array architecture to achieve high array gain while reducing hardware cost. The BS is equipped with $N$ antennas and $N_{\mathrm{RF}}$ RF chains for receiving uplink offloading data, while each device is equipped with $M$ antennas and $M_{\mathrm{RF}}$ RF chains for transmission. Meanwhile, the uplink signals transmitted by  $K$ devices are also exploited for cooperative target detection. Specifically, we adopt a decoding-then-sensing framework \cite{random1, dands1, dands2}, where the BS first decodes the offloaded data from all devices and then performs matched filtering on the received signals using the reconstructed signals. In this manner, user-specific sensing signals can be extracted and utilized for cooperative detection. Meanwhile, time synchronization between the BS and the devices is assumed to be achieved through over-the-air synchronization and calibration mechanisms\footnote{\textcolor{black}{In this work, ideal synchronization is assumed to characterize the achievable performance of the proposed framework, following a common assumption in cooperative sensing and ISAC studies \cite{MECS, syn2, syn}. Although advanced synchronization techniques can provide highly accurate synchronization in practical systems \cite{syna1,syna2}, residual synchronization errors may arise due to hardware imperfections. The robust sensing designs under synchronization uncertainties will be investigated in future work.}} \cite{syna1,syna2, syn}.

Besides executing the computation tasks offloaded by the devices, the BS also performs a DNN-based sensing task, such as post-detection target classification and recognition. 
Specifically, the BS preprocesses the received echo signals using the short-time Fourier transform (STFT) \cite{stft} to generate target-related spectrograms, which are then fed into a pre-trained DNN model for target classification (e.g., pedestrians and vehicles). 
To fully leverage the hierarchical computing architecture and alleviate the computation load on the MEC server, we adopt a cloud–edge collaborative inference scheme, where the MEC server executes the early layers of the DNN and the cloud server processes the remaining layers to produce the final inference results. 
Since the feature dimensionality is significantly compressed at certain intermediate layers (e.g., pooling layers), this split strategy not only reduces the computation load at the MEC server but also mitigates the edge–cloud offloading latency. \textcolor{black}{For clarity, we provide a detailed system flow diagram in Fig. 2.}

\begin{figure}[!t]
	\centering
	\includegraphics[width=2.4in]{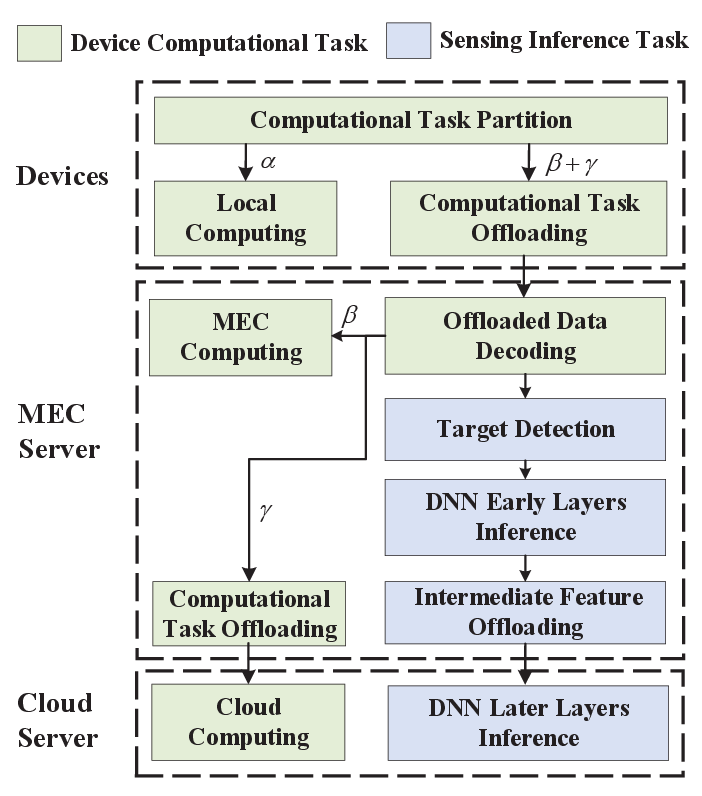}
	\caption{\textcolor{black}{System flow diagram.}}
	\label{fig:flow}
\end{figure}

\subsection{Communication Model}
Let $\mathbf{F}_{\mathrm{D},k}\in\mathbb{C}^{M_{\mathrm{RF}} \times d}$ and  $\mathbf{F}_{\mathrm{RF},k}\in\mathbb{C}^{M  \times  M_{\mathrm{RF}}}$ denote the digital and analog transmit beamforming matrices of device $k$, respectively, where $d$ denotes the number of data streams of each device. Let $\mathbf{V}_{\mathrm{D},k}\in\mathbb{C}^{ N_{\mathrm{RF}} \times d }$ and  $\mathbf{V}_{\mathrm{RF}}\in\mathbb{C}^{N  \times  N_{\mathrm{RF}}}$ represent the digital combining matrices for device $k$ and analog combining matrices at the BS, respectively. Since the analog beamforming matrices $\mathbf{F}_{\mathrm{RF},k}$ and combining matrices $\mathbf{V}_{\mathrm{RF}}$ are implemented via phase shifters, each of their elements is subject to a constant-modulus constraint, i.e., $|[\mathbf{F}_{\mathrm{RF},k}]_{i,j}|=1,   |[\mathbf{V}_{\mathrm{RF}}]_{i,j}|=1, \forall k, i,j$. By defining the hybrid receive combining matrices as $\mathbf{V}_{k}=\mathbf{V}_{\mathrm{RF}}\mathbf{V}_{\mathrm{D},k}$, the received signal after hybrid combining at the BS at time $t$ can be expressed as

\begin{equation} 
\begin{aligned}
\mathbf{r}_{k}(t)= & \underbrace{\mathbf{V}_{k}^H \mathbf{H}_{k} \mathbf{F}_{\mathrm{RF},k} \mathbf{F}_{\mathrm{D},{k}} \mathbf{s}_{k}}_{\text {desired signal }}(t) \\
& +\underbrace{\mathbf{V}_{k}^H  \sum_{i \neq k}^K \mathbf{H}_{i} \mathbf{F}_{\mathrm{RF},i} \mathbf{F}_{\mathrm{D},i} \mathbf{s}_{i}}_{\text { interference }}(t) +\underbrace{\mathbf{V}_{k}^H \mathbf{n}_c}_{\text {noise }}(t) ,
\end{aligned}
\end{equation}
where $\mathbf{s}_{k}(t)\in\mathbb{C}^{d \times 1}$  denotes the data stream of device $k$'s to be offloaded to BS, and $\mathbf{n}_c(t)\sim \mathcal{CN}(0, {\sigma}_c^2\mathbf{I}_{N} )$  represents the additive white Gaussian noise (AWGN) at the BS.  $\mathbf{H}_{k}\in\mathbb{C}^{N \times M}$ denotes the uplink channel matrix from device $k$ to the BS. In this paper, we adopt the mmWave channel model in \cite{channel1,uplink1}. \textcolor{black}{Since sensing and communication signals share the same propagation environment in uplink ISAC systems, the environmental scattering (clutter) components are incorporated into the channel model to characterize the realistic propagation characteristics.} When a sensing target is present, $\mathbf{H}_k$ is modeled as
\begin{equation} 
\mathbf{H}_k=\mathbf{H}^{\text{scatter}}_k+ \mathbf{A}_k,
\end{equation}
where $\mathbf{H}^{\text{scatter}}_k=\sum_{i=1}^{n_{c,k}}\zeta_{k,i}\mathbf{a}(\theta_i)\mathbf{b}^H(\phi_{k,i})$  represents the static environmental clutter components, and $\mathbf{A}_k=e^{j2\pi f^D_{k,0}t}\zeta_{k,0}\mathbf{a}(\theta_0)\mathbf{b}^H(\phi_{k,0})\in\mathbb{C}^{N\times M}$ denotes the target-reflected component. When the target is absent, we have  $\mathbf{H}_k=\mathbf{H}^{\text{scatter}}_k$. Here, $n_{c,k}$ denotes the total number of propagation clusters associated with device $k$. The vectors $\mathbf{a}(\theta_i) = [1, {\rm e}^{j2\pi \delta \sin(\theta_i)}, \ldots, {\rm e}^{j2\pi (N -1) \delta \sin(\theta_i)}]^T \in \mathbb{C}^{N \times 1}$ and $\mathbf{b}(\phi_{k,i}) = [1, {\rm e}^{j2\pi \delta \sin(\phi_{k,i})}, \ldots, {\rm e}^{j2\pi (M -1) \delta \sin(\phi_{k,i})}]^T \in \mathbb{C}^{M \times 1}$ are the receive and transmit steering vectors, respectively, where $\delta$ is  the normalized antenna spacing. The parameters $\theta_i$, $\phi_{k,i}$, and $\zeta_{k,i}$ denote the elevation angle, azimuth angle, and complex path gain of the $i$-th cluster, respectively, \textcolor{black}{which can be obtained through measurement-based characterization or geometry-based modeling of environmental scattering clusters \cite{chmodel1,chmodel2}.} $\theta_0$, $\phi_{k,0}$, and $f^D_{k,0}, \zeta_{k,0}$ correspond to the elevation angle, azimuth angle,  Doppler frequency shift, and reflection coefficient of the device–target–BS link\footnote{We focus on detecting the existence of a target within a given area. Therefore, the target-related parameters $\theta_0$, $\phi_{k,0}$, and $\zeta_{k,0}$ are fixed for analytical tractability \cite{MECS,MECS2}.}.

 Let $\mathbf{F}_{k}=\mathbf{F}_{\mathrm{RF},k}\mathbf{F}_{\mathrm{D},k}$  denote the hybrid transmit beamforming matrices of device $k$. When a capacity-lossless linear minimum mean square error (LMMSE) digital receiver is adopted \cite{wmmse1}, the maximum achievable offloading rate of device $k$ can be expressed as
\begin{equation}
\label{r1}
R_{k} = B\log\det(\mathbf{I}_{N_{\mathrm{RF}}}+\mathbf{V}^H_{\mathrm{RF}}\mathbf{H}_{k}\mathbf{F}_{k}(\mathbf{F}_{k})^H\mathbf{H}^H_{k}\mathbf{V}_{\mathrm{RF}}\mathbf{\Upsilon}^{-1}_{k} ).
\end{equation}
where $B$ denotes the signal bandwidth   and $\mathbf{\Upsilon}_k$ denotes the interference-plus-noise covariance matrix, which is given by
\begin{equation}
\mathbf{\Upsilon}_k=\sum_{i \neq k}^K \mathbf{V}_{\mathrm{RF}}^H \mathbf{H}_{i} \mathbf{F}_{i} \mathbf{F}_{i}^H \mathbf{H}_{i}^H \mathbf{V}_{\mathrm{RF}}+\sigma^2 \mathbf{V}_{\mathrm{RF}}^H \mathbf{V}_{\mathrm{RF}}.
\end{equation}
\subsection{Sensing Model}
For the sensing process, the BS first applies background subtraction or filtering techniques to suppress undesired echo components caused by the clutters\footnote{We assume that static clutter can be suppressed through Doppler-based moving target indication (MTI/MTD) \cite{MTI} and angle-of-arrival (AoA) discrimination using space–time adaptive processing (STAP) \cite{stap}. Signals reflected by moving scatterers cannot be fully eliminated and are therefore treated as sensing targets to enhance environmental awareness.} \cite{wangtcom2022,Andrew2021}. \textcolor{black}{After this preprocessing, the effective sensing signal at the BS can be expressed as \cite{MECS, MECS2}}
\begin{equation}
\label{r1}
\mathbf{r}_s(t)=\sum_{k=1}^K \mathbf{A}_k\mathbf{F}_{\mathrm{RF},k} \mathbf{F}_{\mathrm{D},{k}} \mathbf{s}_k\left(t-\tau_{k,0}\right)+\mathbf{n}_s\left(t\right),
\end{equation}
\textcolor{black}{where $\mathbf{n}_s(t)$  denotes the effective sensing noise after preprocessing.} $\tau_{k,0}$ denote the propagation delay of the device–target–BS path. After decoding the data symbols of each device, we exploit them to perform matched filtering on the received sensing signal $\mathbf{r}_s(t)$\footnote{Since communication links generally require a very low bit-error rate, we assume that the decoded data symbols used for matched filtering suffer from negligible decoding errors. Under this assumption, the decoding process does not noticeably distort the sensing reference, thus enabling reliable sensing performance \cite{dands1,dands2}.}. The output of the matched filter corresponding to device $k$ is given by
\textcolor{black}{
\begin{equation}
\begin{aligned}
\label{eq:yk_def}
\mathbf{Y}_{k}
= &\frac{1}{\sqrt{T}} \int_{0}^{T} \mathbf{r}_s(t)\,\mathbf{s}_k^{H}(t-\tau_k){\rm e}^{-j2\pi ft}\,dt,\\
= &~\frac{1}{\sqrt{T}}\sum_{i=1}^{K}\mathbf{A}_i\mathbf{F}_i
\int_{0}^{T} \mathbf{s}_i(t-\tau_i)\mathbf{s}_k^{H}(t-\tau_k){\rm e}^{-j2\pi ft}\,dt \\
&~+ \frac{1}{\sqrt{T}} \int_{0}^{T} \mathbf{n}_s(t)\,\mathbf{s}_k^{H}(t-\tau_k){\rm e}^{-j2\pi ft}\,dt .
\end{aligned}
\end{equation}}where $T$ denotes the observation interval of the matched filter. Assuming that the data symbols of different devices are temporally white, mutually uncorrelated, and unit-power, the cross-correlation terms with $i\neq k$ average out over a sufficiently long $T$, while the autocorrelation term for $i=k$ satisfies $\int_{0}^{T} \mathbf{s}_k(t-\tau_k)\mathbf{s}_k^{H}(t-\tau_k)\,dt \approx T\mathbf{I}_d.$ \textcolor{black}{By matched-filtering (\ref{eq:yk_def}) using a delayed and
Doppler-shifted version of $\mathbf{s}_k(t)$, the propagation delay $\tau_{k,0}$ and the Doppler frequency $f^D_{k,0}$ can be estimated.} Consequently, the matched filter output can be approximated as \cite{MECS}
\begin{equation}
\label{eq:yk_final}
\mathbf{Y}_k \approx \sqrt{T}\,\mathbf{A}_k\mathbf{F}_k + \mathbf{N}_k,
\end{equation}
where $\mathbf{N}_k \triangleq
\frac{1}{\sqrt{T}} \int_{0}^{T} \mathbf{n}(t)\,\mathbf{s}_k^{H}(t-\tau_k)\,dt$ is the filtered noise matrix.  Stacking all matched–filter outputs and vectorizing yields
\begin{equation}
\mathbf{y}_S 
=
\big[
\mathrm{vec}(\mathbf{Y}_1)^T,\ldots,\mathrm{vec}(\mathbf{Y}_K)^T
\big]^T
\approx 
\hat{\mathbf{g}}+\hat{\mathbf{n}},
\end{equation}
where
\begin{equation}
\begin{aligned}
\hat{\mathbf{g}}
&= \sqrt{T}
\big[
\mathrm{vec}(\mathbf{A}_1\mathbf{F}_1)^T,
\ldots,
\mathrm{vec}(\mathbf{A}_K\mathbf{F}_K)^T
\big]^T,\\
\hat{\mathbf{n}}
&= \big[
\mathrm{vec}(\mathbf{N}_1)^T,
\ldots,
\mathrm{vec}(\mathbf{N}_K)^T
\big]^T .
\end{aligned}
\end{equation}

Thus, the binary hypothesis testing problem can be written as
\begin{equation}
\begin{cases}
\mathcal{H}_0:~ \mathbf{y}_S = \hat{\mathbf{n}}, \\[4pt]
\mathcal{H}_1:~ \mathbf{y}_S  =\hat{\mathbf{g}}+\hat{\mathbf{n}},
\end{cases}
\end{equation}
where $\mathcal{H}_0$ and $\mathcal{H}_1$ denote the absence and presence of the target, respectively.  The probability density functions (PDFs) of $\mathbf{y}_S$ under $\mathcal{H}_0$ and $\mathcal{H}_1$ are given by
\begin{equation}
\label{eq:pdfs}
\begin{cases}
\mathbb{P}(\mathbf{y}_S|\mathcal{H}_0)
=
\dfrac{1}{(\pi\sigma^2)^{D}}
\exp\!\left(-\dfrac{1}{\sigma^2}\mathbf{y}_S^{H}\mathbf{y}_S\right), \\[12pt]
\mathbb{P}(\mathbf{y}_S|\mathcal{H}_1)
=
\dfrac{1}{(\pi\sigma^2)^{D}}
\exp\!\left(-\dfrac{1}{\sigma^2}(\mathbf{y}_S-\hat{\mathbf{g}})^{H}(\mathbf{y}_S-\hat{\mathbf{g}})\right),
\end{cases}
\end{equation}
where $D$ denotes the dimension of $\mathbf{y}_S$. Applying the Neyman–Pearson criterion, the log-likelihood ratio test is written as
\begin{equation}
\ln\frac{\mathbb{P}(\mathbf{y}_S|\mathcal{H}_1)}
{\mathbb{P}(\mathbf{y}_S|\mathcal{H}_0)}=-\frac{1}{\sigma^2}(\mathbf{y}_S-\hat{\mathbf{g}})^{H}(\mathbf{y}_S-\hat{\mathbf{g}})
+\frac{1}{\sigma^2}\mathbf{y}_S^{H}\mathbf{y}_S
\underset{\mathcal{H}_0}{\overset{\mathcal{H}_1}{\gtrless}}
\eta ,
\end{equation}
where $\eta$ denotes the detection threshold corresponding to a prescribed false alarm probability $P_{\rm {fa}}$. The corresponding detection probability is given by
\begin{equation}
P_{D}
=
Q\!\left(
Q^{-1}(P_{\rm {fa}})
-
\frac{\sqrt{2}\|\hat{\mathbf{g}}\|}{\sigma}
\right),
\end{equation}
 \textcolor{black}{where $Q(\cdot)$ denotes the Gaussian $Q$-function, defined as
$
Q(x)=\frac{1}{\sqrt{2\pi}}\int_x^{\infty} e^{-t^2/2}\,dt.
$} To guarantee reliable detection performance, we impose the detection probability requirement $P_D \ge P_D^{\min}$, where $P_D^{\min}$ denotes the minimum required detection probability. By exploiting the monotonicity of $Q(\cdot)$, this constraint can be equivalently expressed as 
\begin{equation}
\|\hat{\mathbf{g}}\|^2
\ge
\frac{\sigma_s^2}{2}
\Big(Q^{-1}(P_{\rm {fa}})-Q^{-1}(P_D^{\min})\Big)^2 \triangleq \Gamma.
\label{eq:g_constraint}
\end{equation}

Since $\|\hat{\mathbf{g}}\|^2=T \sum_{k=1}^{K}\mathrm{Tr}\!\left(\mathbf{A}_k\mathbf{F}_k\mathbf{F}^{H}_k\mathbf{A}_k^{H}\right)$, the detection probability constraint is equivalently written as

\begin{equation}
T \sum_{k=1}^{K}\mathrm{Tr}\!\left(\mathbf{A}_k\mathbf{F}_k\mathbf{F}^{H}_k\mathbf{A}_k^{H}\right)
\ge
\Gamma.
\end{equation}

\subsection{Device computation task Model}

For the computation tasks generated by the devices, we adopt a commonly used task model in MEC research, characterized by $(I_k, c_k)$ \cite{mec3,mec4}, where $I_k$ represents the task input size (in bits) and $c_k$ denotes the 	computation intensity, defined as the number of CPU cycles required to process one bit of data. Accordingly, the total required CPU cycles for device $k$ are given by $I_kc_k$. The task is partitioned and executed in parallel across the local device, the MEC server, and the cloud server with partition ratios $\alpha_k$, $\beta_k$, and $\gamma_k$, respectively, satisfying 
\begin{equation}
\alpha_k+\beta_k+\gamma_k=1,~\forall k\in\mathcal{K}.
\end{equation}

The corresponding computation overhead is expressed as follows.
\subsubsection{Local Execution}
Let $f^\text{L}_k$ denote the local CPU frequency of the device $k$. Then,  the local execution latency and the local computing power consumption can be respectively expressed as 
\begin{equation}
T_k^\text{L}=\frac{\alpha_kI_kc_k}{f^\text{L}_k},~~p^\text{L}_k=\kappa(f^\text{L}_k)^3, ~\forall k\in \mathcal{K}, 
\end{equation}
where $\kappa$ denotes power coefficient related to the hardware architecture.
\subsubsection{MEC Server Execution}
The execution latency at the MEC server consists of the uplink offloading delay $t^\text{O}_k$ from the device to the BS  and the computing delay $t^\text{M}_k$ at the MEC server,  expressed as\footnote{Since the MEC and cloud servers are powered by stable wired infrastructures, their energy supply is generally considered sufficient and is therefore not treated as a limiting factor in this work.}
\begin{equation}\label{t2}
T_k^\text{E}=t^\text{O}_k+t^\text{M}_k=\frac{(\beta_k+\gamma_k)I_k}{R_{k}}+\frac{\beta_kI_kc_k}{f^\text{M}_k}, ~\forall k\in \mathcal{K}.
\end{equation}
where $f^\text{M}_k$ represents the CPU frequency allocated to device $k$ at the MEC server.
\subsubsection{Cloud Server Execution}
The cloud execution latency consists of the device–MEC offloading delay $t^\text{O}_k$, the MEC–cloud transmission delay $t^\text{B}_k$, and the computing delay  $t^\text{C}_k$ at the cloud server, expressed as
\begin{equation}\label{t3}
T_k^\text{C}=t_k^\text{O}+t_k^\text{B}+t_k^\text{C}=\frac{(\beta_k+\gamma_k)I_k}{R_{k}}+\frac{\gamma_kI_k}{r_b}+\frac{\gamma_kI_kc_k}{f^\text{C}_k}, ~\forall k\in \mathcal{K}.
\end{equation}
where $r_b$ denotes the optical fiber transmission rate between the MEC server and the cloud server, and $f_k^\text{C}$ represents the computing frequency allocated to device $k$ at the cloud server. Since the computation results are typically much smaller than the original computation data, the feedback latency from the cloud and MEC servers is neglected, as in \cite{mec3,mec4}.

 Accordingly, for the computation task generated by device $k$, the overall task completion time  is determined by the slowest computing branch among the local execution, MEC execution, and cloud execution, and is therefore given by
\begin{equation}
T_k^{\mathrm{tot}}
= \max\!\left\{\,T_k^{\mathrm{L}},~T_k^{\mathrm{E}},~T_k^{\mathrm{C}}\,\right\},
\quad \forall k \in \mathcal{K}.
\end{equation}

\subsection{\textcolor{black}{Sensing Inference Task Model}}
\textcolor{black}{Besides partitioning device computation tasks among the local devices, MEC server, and cloud server, the considered cloud–edge–device architecture also enables collaborative processing of computation-intensive sensing inference tasks. To this end, a DNN split-inference strategy is introduced to distribute sensing inference workloads between the MEC and cloud servers. The selected split point determines not only the DNN layers executed at the MEC and cloud servers, which affects their computation workloads, but also the size of intermediate features transmitted through the MEC-cloud link. Consequently, it influences the allocation of computation and communication resources and the optimal task partitioning decisions in the cloud–edge–device architecture.}

\textcolor{black}{After target detection, the BS employs a DNN-based model for target classification. Specifically, the echo signals associated with the detected targets are first transformed into time–frequency spectrograms via the STFT\footnote{\textcolor{black}{Radar preprocessing operations, such as STFT and GMTI, may introduce non-negligible computational overhead. In this work, they are regarded as conventional front-end processing modules completed before DNN inference, whose complexity depends on radar configurations and hardware implementations \cite{Radarpre1}, rather than the resource allocation considered herein. Therefore, their latency is modeled as a fixed sensing overhead independent of the optimization variables in our framework \cite{dnnmec3,dnniscc3}. The end-to-end joint optimization of radar signal preprocessing and DNN-based inference is left for future work.}}\cite{stft}. These spectrograms are then fed into a pre-trained DNN to extract semantic features and classify the targets into categories (e.g., pedestrians and vehicles) \cite{Alexnet1}.} To reduce the computational burden on the MEC server, we adopt a split-inference strategy, where the early layers of the DNN are executed at the MEC server and the remaining layers are processed at the cloud server for collaborative inference. We use floating-point operations (FLOPs) to characterize the computation resources required by different DNN layers. For clarity, the set of pre-trained DNN layers is denoted by
$\mathcal{L}_=\{1,\ldots,L\}$, where the FLOPs required by the $l$-th layer are denoted by $z(l)$, which depends on its layer type $\Xi(l)$.  The main computation load of the DNN originates from  fully connected (FC), convolutional (CV), and max-pooling (MP) layers, with the FLOPs of each layer type given by \cite{dnnmec1}
\begin{align}
\begin{split}
z(l)= \left \{
\begin{array}{ll}
(2a(l-1)-1)a(l), & \Xi(l)=\text{FC},\\
(2b(l-1)c(l)^2-1)d(l) e(l)b(l),                  & \Xi(l)=\text{CV},\\
b(l) c^2(l)d(l)e(l),                    & \Xi(l)=\text{MP},
\end{array}
\right.
\end{split}
\end{align}
where $a(l)$, $b(l)$, $c(l)$, $d(l)$, and $e(l)$ respectively represent the number of neurons in the $l$-th layer, the number of channels, the filter size, and the height and width of the corresponding output feature map.  The inference latency of the DNN-based sensing task, denoted by $T_0^\text{I}$, consists of the  inference delay at the MEC server $t_0^\text{M}$, the transmission delay of intermediate features from the MEC server to the cloud server  $t_0^\text{B}$, and the inference delay at the cloud server $t_0^\text{C}$. Let $l_s$ denote the splitting point.  Then, $T_0^\text{I}$ can be expressed as
\begin{equation}\label{t6}
T_0^\text{I}=t_0^\text{M}+t_0^\text{B}+t_0^\text{C}=\frac{\sum_{l = 1}^{l_s}z(l)}{\rho^\text{M}_0f^\text{M}_0}+\frac{q(l_s)}{r_b}+\frac{ \sum_{l = l_s+1}^{L}z(l)}{\rho^\text{C}_0f^\text{C}_0},
\end{equation}
where $\rho^\text{M}_0$  and $\rho^\text{C}_0$ denote the FLOPs per CPU cycle of the MEC server and cloud server; \textcolor{black}{$q(l_s)$ denotes the output data size (in bits) of the $l_s$-th layer, which is uniquely determined by $\Xi(l_s)$,} and  $f^\text{M}_0$ and $f^\text{C}_0$  denote the allocated CPU frequencies for the sensing task at the MEC and cloud servers, respectively. It is observed from (\ref{t6}) that the choice of the splitting point $l_s$ not only determines how the computation burden of the sensing task is distributed between the MEC server and the cloud server, but also affects the size of the intermediate features that need to be transmitted. By appropriately selecting $l_s$, the inference latency of the sensing task can be effectively reduced.

\subsection{Problem Formulation}
\textcolor{black}{The primary objective of this work is to minimize the execution latency of device computation tasks in a cloud--edge--device collaborative computing architecture. However, the introduction of sensing functionality couples the communication and computation processes at both the physical and resource allocation layers, leading to intricate tradeoffs that impact overall task execution latency. These interdependencies manifest in two key aspects.
}

\begin{itemize}
	\item \textcolor{black}{At the \textbf{physical layer}, spatial coupling arises due to signal reuse in the considered ISCC architecture, where the BS performs passive sensing by reusing uplink computation offloading signals.  This leads to an  tradeoff in hybrid beamforming design: to reduce transmission latency, beamforming should focus energy toward the BS to maximize offloading rates, whereas satisfying detection probability constraints requires sufficient beamforming gain toward target directions. Consequently, hybrid beamforming must balance between communication directivity and sensing coverage in the spatial domain.}

	\item \textcolor{black}{At the \textbf{computation layer}, resource coupling arises due to shared computing resources. Unlike conventional MEC systems, the MEC server in the proposed architecture serves as a common pool for both device computation offloading and DNN-based sensing inference tasks. Through split inference, sensing tasks consume a portion of edge and cloud computing resources determined by the DNN splitting point $l_s$. While executing more sensing layers at the MEC server reduces sensing inference latency, it limits computation resources for device tasks, potentially increasing overall execution latency. Consequently, the competition for limited resources requires joint optimization of task offloading ratios and sensing model partitioning strategies.
	}
	
\end{itemize}

\textcolor{black}{Based on the above analysis, we model sensing performance requirements, including detection probability and inference latency, as system constraints and formulate a joint optimization problem to minimize the average execution latency of computation tasks across all devices\footnote{\textcolor{black}{In this work, the DNN model is assumed to be pre-trained and fixed, and split inference only partitions the model between the MEC and cloud servers without changing the model architecture or parameters. \textcolor{black}{Under sufficiently accurate intermediate-feature transmission, the classification accuracy remains unchanged with different split points. When feature quantization or compression is considered, different split points may result in different intermediate-feature payloads and accuracy degradation, whose impact is further evaluated in the simulation results.}}}}. Specifically, by jointly optimizing the task partition ratios $\boldsymbol{\alpha}$, $\boldsymbol{\beta}$, and $\boldsymbol{\gamma}$, multi-tier computation resource allocation $\bm{f^\text{L}}, \bm{f^\text{M}}, \bm{f^\text{C}}$, hybrid beamforming matrices $\{\mathbf{F}_k, \mathbf{V}_k\}$, and the DNN splitting point $l_s$, the proposed framework minimizes the average latency of all device tasks while ensuring the required sensing quality\footnote{\textcolor{black}{The average computation latency is adopted to capture overall system efficiency in multi-device scenarios and is widely used in MEC studies \cite{mec3,peng2024tcom}. The proposed framework can also be extended to fairness-oriented objectives (e.g., min–max latency) by introducing an auxiliary worst-case latency variable, without changing the subproblem structure.}}. The resulting optimization problem is formulated as follows:

\begin{align} 
\label{prob29}
\min_{\substack{ \bm{f}^\text{L}, \bm{f}^\text{M},\bm{f}^\text{C}, l_s\\{\mathbf{F}_k, \mathbf{V}_k}, \bm{\alpha}, \bm{\beta},\bm{\gamma}}}   & ~\frac{1}{K}\sum_{k=1}^{K}T_k^{\mathrm{tot}}~\\\text{s.t.}~~~
& \sum_{k=0}^{K}f^\text{M}_k\leq f^\text{M}_\text{th}, ~~\sum_{k=0}^{K}f^\text{C}_k\leq f^\text{C}_\text{th}, \tag{\ref{prob29}a}\\
& 0\leq{f}_k^\text{L}\leq f^\text{L}_\text{th},~~~p^\text{L}_k\leq P^\text{L}_\text{th}, ~~\forall k\in \mathcal{K}, \tag{\ref{prob29}b}\\
& |\mathbf{F}_k|^2_F\leq P^\text{O}_\text{th}, ~~~\forall k \in \mathcal{K}, \tag{\ref{prob29}c}\\
&|[\mathbf{F}_{\mathrm{RF},k}]_{i,j}|=1,  \forall k\in \mathcal{K},i,j ,  \tag{\ref{prob29}d}\\
&|[\mathbf{V}_{\mathrm{RF}}]_{i,j}|=1,~~\forall i,j , \tag{\ref{prob29}e}\\
&T\sum_{k=1}^{K}\mathrm{Tr}\!\left(\mathbf{A}_k\mathbf{F}_k\mathbf{F}^{H}_k\mathbf{A}_k^{H}\right)\ge\Gamma, \tag{\ref{prob29}f}\\
& \alpha_k+\beta_k+\gamma_k=1, ~\forall k\in \mathcal{K}, \tag{\ref{prob29}g}\\
&0 \leq \alpha_k,\beta_k,\gamma_k\leq 1, ~~\forall k\in \mathcal{K},\tag{\ref{prob29}h}\\
&T_0^\text{I}\leq T_\text{th}^\text{I}, \tag{\ref{prob29}i}\\&l_s\in \{1,\cdots,L\} ,\tag{\ref{prob29}j}
\end{align}
where $f^\text{L}_\text{th}$, $f^\text{M}_\text{th}$, and $f^\text{C}_\text{th}$  denote the computational capacities of device 
$k$, the MEC server, and the cloud server, respectively; $P^\text{L}_\text{th}$ and $P^\text{O}_\text{th}$ represent the power budgets for local computation and signal transmission at the device, respectively; and $T_\text{th}^\text{I}$ denotes the maximum tolerable latency of the sensing task. The complicated coupling among variables in the objective function, together with the constant-modulus constraints  (\ref{prob29}d)--(\ref{prob29}e) and the  non-convex detection probability constraint (\ref{prob29}f), renders problem (\ref{prob29}) highly non-convex and computationally intractable. Moreover, the incorporation of hybrid beamforming and the integer-valued DNN splitting point variable further  exacerbates the optimization challenge.

\section{Proposed Algorithm}
In this section, we propose an alternating optimization (AO) algorithm to efficiently solve problem (\ref{prob29}), which integrates closed-form derivation, WMMSE reformulation, and manifold optimization techniques. Specifically, by exploiting the block-wise structure of the problem (\ref{prob29}), we decompose it into the following three subproblems:
\begin{itemize}
	\item \textbf{Computation resource allocation and DNN splitting point optimization:}  For the joint optimization of $\{\bm{f^\text{L}}, \bm{f^\text{M}}, \bm{f^\text{C}}\}$ and  $l_s$, we leverage a hierarchical decoupling strategy and develop a two-layer optimization framework. In the inner layer, given a fixed splitting point $l_s$, we derive the closed-form solutions of $\{\bm{f^\text{L}}, \bm{f^\text{M}}, \bm{f^\text{C}}\}$. In the outer layer, we search over the candidate $l_s$ and select the one that yields the minimum execution latency.
	\item \textbf{Hybrid beamforming and combining matrices optimization:} For the optimization subproblem involving $\{\mathbf{F}_k\}_{\forall k}, \{\mathbf{V}_k\}_{\forall k}$, we first transform the original formulation into a tractable form via the WMMSE approach. Then, we jointly optimize the digital beamformers/combiners and their analog counterparts using the SCA technique, together with a manifold optimization algorithm augmented with penalty terms.
	\item \textbf{Task partition optimization:} For the  subproblem with respect to $\{\bm{\alpha}, \bm{\beta},\bm{\gamma}\}$, by exploiting the equalization principle inherent in the min-max objective, we derive the optimal closed-form solutions for the task partition ratios.
\end{itemize}

In the following, we present the implementation details of each part.

\subsection{Computation Resource Allocation and DNN splitting point Optimization}
In this subsection, we optimize computation resource allocation $\{\bm{f^\text{L}}, \bm{f^\text{M}}, \bm{f^\text{C}}\}$ and DNN splitting point $l_s$ while fixing $\left\{\{\mathbf{F}_k\}_{\forall k}, \{\mathbf{V}_k\}_{\forall k},\bm{\alpha}, \bm{\beta},\bm{\gamma}\right\}$. {Since $l_s$ represents the discrete DNN layer partition between the MEC and cloud servers, the optimization problem involves both discrete split decisions and continuous resource allocation variables. Therefore, we reformulate this subproblem into a two-layer optimization framework to separate the discrete split decision from the continuous resource allocation variables, thereby avoiding the mixed-integer optimization and facilitating the derivation of closed-form solutions for the continuous variables.} The inner-layer optimization problem is expressed as
\begin{align} 
\label{prob9}
\left(\text{Inner}\right):\min_{ \bm{f}^\text{L}, \bm{f}^\text{M},\bm{f}^\text{C},\hat{t}_k^{\text{tot}}} &  ~~\frac{1}{K}\sum_{k=1}^{K}~\hat{t}_k^{\mathrm{tot}} \\
\text{s.t.}\quad~&  \hat{t}_k^{\mathrm{tot}} \geq \frac{\alpha_kI_kc_k}{f^\text{L}_k},~~\forall k\in \mathcal{K},~\tag{\ref{prob9}a}\\& \hat{t}_k^{\mathrm{tot}} \geq t^\text{O}_k+\frac{\beta_kI_kc_k}{f^\text{M}_k},~~\forall k\in \mathcal{K},\tag{\ref{prob9}b}\\&\hat{t}_k^{\mathrm{tot}} \geq t_k^\text{O}+t_k^\text{B}+\frac{\gamma_kI_kc_k}{f^\text{C}_k},~~\forall k\in \mathcal{K},\tag{\ref{prob9}c}\\
&\text{(\ref{prob29}\text{a})},\text{(\ref{prob29}\text{b})},\text{(\ref{prob29}\text{i})}, \notag
\end{align}
where $t^\text{O}_k$ and $t_k^\text{B}$ are constant latency terms independent of the optimization variables $\{\bm{f^\text{L}}, \bm{f^\text{M}}, \bm{f^\text{C}}\}$, and $\hat{t}_k^{\text{tot}}$ are introduced auxiliary variables. The outer-layer optimization problem is then given by
\begin{align}
\hspace*{-3cm} 
\label{prob8}
\left(\text{Outer}\right):  \min_{l_s} &  ~~\frac{1}{K}\sum_{k=1}^{K}~T_k^{\mathrm{tot}}\\
\text{s.t.}~& ~l_s\in \{1,\cdots,L\}.
\end{align}

Considering that the solution space of the outer-layer optimization problem is discrete and finite, we can obtain the optimal solution to this subproblem by performing a one-dimensional search over $l_s$, and solving the inner-layer problem for each  $l_s$. To further reduce the computational complexity incurred by repeatedly solving the inner-layer problem, we derive the optimal closed-form solutions of $\{\bm{f^\text{L}}, \bm{f^\text{M}}, \bm{f^\text{C}}\}$. Specifically,  since each device seeks to minimize its local execution latency by utilizing the maximum feasible local computing frequency under its power and capacity constraints, it is straightforward to show that the optimal local computing frequency of device $k$ is given by
\begin{equation}\label{c0}
f^{\text{L}*}_k = \min\left\{f^\text{L}_\text{th},~ \sqrt[3]{\frac{P^\text{L}_\text{th}}{\kappa}}\right\},~\forall k\in \mathcal{K}.
\end{equation}

Then, we analytically derive the closed-form solutions of $\{\bm{f^\text{L}}, \bm{f^\text{M}}, \bm{f^\text{C}}\}$ by exploiting KKT conditions \cite{2004Convex}. Due to the complicated constraint structure of problem (\ref{prob9}), we develop a two-stage Lagrangian framework to facilitate this derivation. In particular, the partial Lagrangian of (\ref{prob9}), associated with constraints (\ref{prob9}a)–(\ref{prob9}c) while keeping the remaining constraints in the feasible set, is given by
\begin{equation} 
	\begin{aligned} 
\label{prob10}
&~~\mathcal{L}_p= ~~\frac{1}{K}\sum_{k=1}^{K}~\hat{t}_k^{\mathrm{tot}} + \lambda^{\text{L}}_k\left(\frac{\alpha_kI_kc_k}{f^\text{L}_k}-\hat{t}_k^{\mathrm{tot}}\right)+ \\&\lambda^{\text{M}}_k\left(t^\text{O}_k+\frac{\beta_kI_kc_k}{f^\text{M}_k}-\hat{t}_k^{\mathrm{tot}}\right)+\lambda^{\text{C}}_k\left(t_k^\text{O}+t_k^\text{B}+\frac{\gamma_kI_kc_k}{f^\text{C}_k}-\hat{t}_k^{\mathrm{tot}}\right) ,
\end{aligned}
\end{equation}
where $\bm{\lambda}=\{\lambda^{\text{L}}_k,\lambda^{\text{M}}_k,\lambda^{\text{C}}_k\}_{\forall k\in \mathcal{K}}$ are the Lagrange multipliers. Accordingly, the dual problem of problem (\ref{prob9}) is formulated as
\begin{align} 
\label{prob11}
\max_{ \bm{\lambda}} &  ~~\widetilde{\mathcal{L}}_p(\bm{\lambda}),
\end{align}
where $\widetilde{\mathcal{L}}_p$ is the dual function given by
\begin{align} 
\label{prob12}
\widetilde{\mathcal{L}}_p(\bm{\lambda})= \min_{ \bm{f}^\text{C} \bm{f}^\text{M},\hat{t}_k^{\text{tot}}}  &~~{\mathcal{L}}_p, ~~\notag \\ \text{s.t.}~~&~ \text{(\ref{prob29}\text{a})},\text{(\ref{prob29}\text{i})}. 
\end{align}

Since problem (\ref{prob9}) is convex, solving the dual problem (\ref{prob12}) is equivalent to solving the primal problem (\ref{prob9}). By iteratively maximizing the dual function in (\ref{prob12}) and updating the associated Lagrange multipliers, problem (\ref{prob9}) can be optimally solved. \textcolor{black}{Denoting the iteration index as $it$, next, we present the derivation of the closed-form solutions for $\bm{f}^\text{M}$ and $\bm{f}^\text{C}$, as well as the updating rule of $\bm{\lambda}$.}
\subsubsection{Updating $\bm{f}^\text{M}$ and $\bm{f}^\text{C}$}
For notational convenience, we define $D_1=\frac{\sum_{l = 1}^{l_s}z(l)}{\rho^\text{M}_0}$ and $D_2=\frac{ \sum_{l = l_s+1}^{L}z(l)}{\rho^\text{C}_0}$.  Accordingly, the Lagrangian function of (\ref{prob12}) with given $\bm{\lambda}^{(it)}$ can be formulated as
\begin{equation} 
\begin{aligned} 
\label{prob13}
&~~\mathcal{L}_c= ~~{\mathcal{L}}_p + \Lambda^{\text{M}}\left(\sum_{k=0}^{K}f^\text{M}_k-f^\text{M}_\text{th}\right)+ \\&\Lambda^{\text{C}}\left(\sum_{k=0}^{K}f^\text{C}_k-f^\text{C}_\text{th}\right)+\mu\left(\frac{D_1}{f^\text{M}_0}+\frac{D_2}{f^\text{C}_0}+t_0^\text{B}-T_\text{th}^\text{I}\right) ,
\end{aligned}
\end{equation}
where $\Lambda^{\text{M}}$, $\Lambda^{\text{C}}$, and $\mu$ are the Lagrange multipliers. When the split strategy is fixed, $D_1$, $D_2$, and $t_0^\text{B}$ are constants. By applying the KKT conditions $\tfrac{\partial \mathcal{L}_c}{\partial f^\text{M}_k} =0$ and $\tfrac{\partial \mathcal{L}_c}{\partial f^\text{C}_k} =0$, the optimal closed-form solutions of $\bm{f}^\text{M}$ and $\bm{f}^\text{C}$ are derived as follows
\begin{align}
f^{\text{M}*}_k &= \sqrt{\frac{\lambda^{\text{M}(it)}_k\beta_kI_kc_k}{\Lambda^{\text{M}}}},~~f^{\text{C}*}_k = \sqrt{\frac{\lambda^{\text{C}(it)}_k\gamma_kI_kc_k}{\Lambda^{\text{C}}}},\forall k\in \mathcal{K}, \label{c1}\\
f^{\text{M}*}_0 &= \sqrt{\frac{\mu D_1}{\Lambda^{\text{M}}}},~~\quad\quad f^{\text{C}*}_0 = \sqrt{\frac{\mu D_2}{\Lambda^{\text{C}}}}.\label{c2}
\end{align}

Since the computation resource allocation for sensing tasks, $f^{\text{M}*}_0$ and $f^{\text{C}*}_0$ cannot be zero or infinite, it follows from (\ref{c1}) and (\ref{c2}) that $\Lambda^{\text{M}},\Lambda^{\text{C}},\mu\neq0 $. According to the complementary slackness condition, constraints (\ref{prob29}\text{a}) and (\ref{prob29}\text{i}) must hold with equality at the optimum. Therefore, by substituting (\ref{c1}) and (\ref{c2}) into constraint (\ref{prob29}a), we obtain
\begin{equation} 
\begin{aligned}\label{c3}
\Lambda^{\text{M}}(\mu)=\left(\frac{\sum_{k=1}^{K}\sqrt{\lambda^{\text{M}(it)}_k\beta_kI_kc_k}+\sqrt{\mu D_1}}{f^\text{C}_\text{th}}\right)^2,\\
\Lambda^{\text{C}}(\mu)=\left(\frac{\sum_{k=1}^{K}\sqrt{\lambda^{\text{C}(it)}_k\gamma_kI_kc_k}+\sqrt{\mu D_2}}{f^\text{C}_\text{th}}\right)^2.
\end{aligned}
\end{equation}

By substituting (\ref{c3}) into (\ref{prob29}\text{i}), we obtain 
\textcolor{black}{\begin{equation}\label{c5}
 \sqrt{\frac{\Lambda^{\text{M}}(\mu)D_1}{\mu}}+ \sqrt{\frac{\Lambda^{\text{M}}(\mu)D_2}{\mu}}=T_\text{th}^\text{I}-t_0^\text{B}.
\end{equation} }
 
We observe that the only remaining unknown variable in (\ref{c5}) is $\mu$, which can be obtained via a bisection search. Once $\mu$ is determined, the corresponding values of $\Lambda^{\text{M}}$ and $\Lambda^{\text{C}}$, as well as the   $\bm{f}^{\text{M}*}$ and $\bm{f}^{\text{C}*}$ can be accordingly obtained.

\subsubsection{Updating $\bm{\lambda}$}
After updating $\bm{f}^{\text{M}}$ and $\bm{f}^{\text{C}}$, the Lagrange multiplier can be updated using the gradient descent method (GDM) \cite{2004Convex}, i.e., 
\begin{equation}
\begin{aligned}
\label{c4}
\lambda^{\text{L}(it+1)}_k&=\lambda^{\text{M}(it)}_k+\eta\left(T_k^{\mathrm{L}(it)}-\hat{t}_k^{\mathrm{tot}(it)}\right),\\
\lambda^{\text{M}(it+1)}_k&=\lambda^{\text{M}(it)}_k+\eta\left(T_k^{\mathrm{M}(it)}-\hat{t}_k^{\mathrm{tot}(it)}\right),\\\lambda^{\text{C}(it+1)}_k&=\lambda^{\text{C}(it)}_k+\eta\left(T_k^{\mathrm{C}(it)}-\hat{t}_k^{\mathrm{tot}(it)}\right),\\
\end{aligned}
\end{equation}
where $\eta$ denotes the step size of the GDM, $T_k^{\mathrm{L}(it)}$, $T_k^{\mathrm{M}(it)}$, and $T_k^{\mathrm{C}(it)}$ represent the local, MEC server, and cloud execution latencies, respectively, evaluated at the computing frequencies obtained in the $it$-th iteration from (\ref{c1}), (\ref{c2}), and (\ref{c3}). 

\textbf{Summary:} To solve the subproblem with respect to $\{\bm{f^\text{L}}, \bm{f^\text{M}}, \bm{f^\text{C}}\}$ and $l_s$, we decompose it into a two-layer structure according to the types of variables. For the outer-layer optimization involving the discrete variable $l_s$, a one-dimensional search is performed to determine the optimal splitting point. For the inner-layer  optimization problem regarding computation resource allocation, we employ  a two-stage Lagrangian framework to derive the optimal closed-form solutions, and provide the corresponding Lagrange multiplier update rules. For clarity, the proposed two-layer optimization algorithm is summarized in Algorithm 1.
{\renewcommand{\baselinestretch}{1}
\begin{algorithm}[t]
	\caption{Two-layer optimization algorithm for  $\{\boldsymbol{f}^\text{L}, \boldsymbol{f}^\text{M}, \boldsymbol{f}^\text{C}\}$ and $l_s$}
	\begin{algorithmic}[1]
		\renewcommand{\algorithmicrequire}{\textbf{Input:}}
		\renewcommand{\algorithmicensure}{\textbf{Output:}}
		\Require Hybrid beamforming and combining matrices $\{\mathbf{F}_k\}_{\forall k}$, $\{\mathbf{V}_k\}_{\forall k}$, and task partition $\boldsymbol{\alpha}, \boldsymbol{\beta}, \boldsymbol{\gamma}$.
		\State Initialize $\boldsymbol{\lambda}^0$ and $it=0$.
		\For{$l_s = 1:L$}
		\Repeat
		\State Obtain the optimal local computing frequency $\boldsymbol{f}^\text{L}$ according to (\ref{c0}).
		\State Obtain the optimal MEC server and cloud server computation resource allocation $\boldsymbol{f}^\text{M}$ and $\boldsymbol{f}^\text{C}$ according to (\ref{c1}) and (\ref{c2}).
		\State Update $\boldsymbol{\lambda}^{(it+1)}$ according to (\ref{c4}).
		\State Update $it=it+1$.
		\Until{the objective value of (\ref{prob9}) converges.}
		\State Record the objective value for the current $l_s$.
		\EndFor
		\Ensure Computation resource allocation $\{\boldsymbol{f}^\text{L}, \boldsymbol{f}^\text{M}, \boldsymbol{f}^\text{C}\}$ and DNN splitting point $l_s$ that yields the minimum objective value.
	\end{algorithmic}
\end{algorithm}}

\subsection{Hybrid Beamforming and Combining  Optimization}
Next, we optimize the hybrid beamforming matrices $\{\mathbf{F}_k\}_{\forall k}$ and combining matrices $\{\mathbf{V}_k\}_{\forall k}$ under given $\{\boldsymbol{\alpha}, \boldsymbol{\beta}, \boldsymbol{\gamma},\boldsymbol{f}^\text{L}, \boldsymbol{f}^\text{M}, \boldsymbol{f}^\text{C},l_s\}$. The corresponding subproblem can be reformulated as
\begin{align} 
\label{prob15}
\min_{ \mathbf{F}_k, \mathbf{V}_k} &  ~~\frac{1}{K}\sum_{k=1}^{K}~\frac{(\beta_k+\gamma_k)I_k}{R_{k}}, \quad  ~~
\text{(\ref{prob29}\text{c})}-\text{(\ref{prob29}\text{f})}. 
\end{align}

The objective of (\ref{prob15}) is highly non-convex, since it involves a sum of terms with log-determinant expressions in the denominator. To handle this issue in a tractable manner, we adopt the SCA technique to linearly approximate it at a given point $\widehat{R}_k$, i.e.,
{\color{black}{\begin{align}
\frac{(\beta_k+\gamma_k)I_k}{R_k}
&\approx
\frac{(\beta_k+\gamma_k)I_k}{\hat{R}_k}
-\frac{(\beta_k+\gamma_k)I_k}{\hat{R}_k^2}
(R_k-\hat{R}_k) \nonumber\\
&=
-\frac{(\beta_k+\gamma_k)I_k}{\hat{R}_k^2}R_k
+\frac{2(\beta_k+\gamma_k)I_k}{\hat{R}_k}.
\end{align}}
}
By defining $\omega_k=\frac{(\beta_k+\gamma_k)I_k}{\widehat{R}^2_k}$ and omitting the constant terms, the subproblem (\ref{prob15}) can be transformed into a weighted sum-rate maximization problem:
\begin{equation} 
\label{prob16}
\max_{ \mathbf{F}_k, \mathbf{V}_k}  ~~\sum_{k=1}^{K}~\omega_k{R_{k}},\quad  \text{s.t.}~\text{(\ref{prob29}\text{c})}-\text{(\ref{prob29}\text{f})}. 
\end{equation}

Subsequently, by leveraging the equivalence between the WMMSE formulation and the sum-rate expression \cite{peng2024tcom}, the objective function of problem (\ref{prob16}) can be rewritten as
\begin{align} 
\label{prob17}
\min_{\mathbf{W}_k,\mathbf{F}_k,\mathbf{V}_k} &  ~~\sum_{k=1}^{K}\omega_k(\text{Tr}(\mathbf{W}_k\mathbf{E}_k)-\log\det(\mathbf{W}_k)),
\end{align}
where $\{\mathbf{W}_k\}_{\forall k}$ are the introduced auxiliary  variables, and $\mathbf{E}_k$ denotes the MSE matrix of device $k$  associated with $\mathbf{F}_k$ and $\mathbf{V}_k$, which is given by
\begin{equation}
\begin{aligned}
\label{mse}
\mathbf{E}_k&=\mathbb{E}\left[\left( \mathbf{r}_k(t)-\mathbf{s}_k(t)\right)\left( \mathbf{r}_k(t)-\mathbf{s}_k(t)\right)^{{H}}\right]\\
&=\mathbf{I}_d -\mathbf{V}^H_k\mathbf{H}_k\mathbf{F}_k-\mathbf{F}^H_k\mathbf{H}^H_k\mathbf{V}_k\\&\quad+{\sum_{i=1}^K\mathbf{V}^H_k\mathbf{H}_{i}\mathbf{F}_{i}(\mathbf{F}_{i})^H\mathbf{H}_{i}^H\mathbf{V}_k}+{\sigma}^2\mathbf{V}_k^H\mathbf{V}_k.
\end{aligned}
\end{equation}

Then, by exploiting the tractable structure of (\ref{prob17}), we employ the block coordinate descent (BCD) technique to alternately update the digital and analog transmit beamforming matrices $\mathbf{F}_{\mathrm{D},k}$ and $\mathbf{F}_{\mathrm{RF},k}$, the digital and analog combining matrices  $\mathbf{V}_{\mathrm{D},k}$ and $\mathbf{V}_{\mathrm{RF}}$, as well as the  auxiliary  variables $\mathbf{W}_k$, thereby solving problem (\ref{prob16}).
\subsubsection{Updating $\mathbf{F}_{\mathrm{D},k}$} With given $\{\mathbf{F}_{\mathrm{RF},k},\mathbf{V}_{\mathrm{D},k},\mathbf{V}_{\mathrm{RF}},\mathbf{W}_k\}$, the optimization subproblem with respect to $\mathbf{F}_{\mathrm{D},k}$ is reformulated as
\begin{align} 
\label{prob18}
\min_{\mathbf{F}_{\mathrm{D},k}} &  ~~\sum_{k=1}^{K}\omega_k\text{Tr}(\mathbf{W}_k\mathbf{E}_k)\\\text{s.t.}&\quad
\text{(\ref{prob29}\text{c})},\text{(\ref{prob29}\text{f})}. \notag
\end{align}

It can be observed that (\ref{prob29}f) is the only non-convex component in problem (\ref{prob18}). To address this issue, we adopt the SCA technique to linearize it at a given point $\widehat{\mathbf{F}}_{\mathrm{D},k}$ as
\begin{equation}
\begin{aligned} \label{prob19}
& \sum_{k=1}^K\left[2 \text{Re}\left\{\operatorname{Tr}\left(\mathbf{A}_k \mathbf{F}_{\mathrm{RF},k}{\mathbf{F}}_{\mathrm{D},k}\widehat{\mathbf{F}}^H_{\mathrm{D},k}\mathbf{F}_{\mathrm{RF},k}^H \mathbf{A}_k^H\right)\right\}\right. \\
& \quad \left.-\operatorname{Tr}\left(\mathbf{A}_k \mathbf{F}_{\mathrm{RF},k}\widehat{\mathbf{F}}_{\mathrm{D},k}\widehat{\mathbf{F}}^H_{\mathrm{D},k}\mathbf{F}_{\mathrm{RF},k}^H \mathbf{A}_k^H\right)\right] \geq \Gamma  / T.
\end{aligned}
\end{equation}

After replacing constraint (\ref{prob29}f) with (\ref{prob19}), problem (\ref{prob18}) is transformed into a convex problem and can be efficiently solved using the interior-point method.
\subsubsection{Updating $\mathbf{F}_{\mathrm{RF},k}$}
When  $\{\mathbf{F}_{\mathrm{D},k},\mathbf{V}_{\mathrm{D},k},\mathbf{V}_{\mathrm{RF}},\mathbf{W}_k\}$ are fixed, the optimization of the analog beamforming matrices ${\mathbf{F}}_{\mathrm{RF},k}$ is subject to the constant-modulus constraint, as well as the power and detection probability constraints. To improve tractability, we incorporate constraints (\ref{prob29}c) and (\ref{prob29}f) into the objective function as penalty terms, and reformulate the subproblem as
\begin{align} 
\label{prob20}
& \min_{\mathbf{F}_{\mathrm{RF},k}} \sum_{k=1}^{K}\omega_k\text{Tr}(\mathbf{W}_k\mathbf{E}_k) \notag +\sum_{k=1}^{K}\upsilon_{1,k}\left(\left[ |\mathbf{F}_{\mathrm{RF},k}\mathbf{F}_{\mathrm{D},k}|^2_F- P^\text{O}_\text{th}\right]^+\right)^2\\&\quad+\upsilon_2\left(\left[
\frac{\Gamma}{T} - \sum_{k=1}^{K}\mathrm{Tr}\!\left(\mathbf{A}_k\mathbf{F}_{\mathrm{RF},k}\mathbf{F}_{\mathrm{D},k}\mathbf{F}^H_{\mathrm{D},k}\mathbf{F}^H_{\mathrm{RF},k}\mathbf{A}_k^{H}\right)\right]^+\right)^2\\&~~\text{s.t.}\quad
\text{(\ref{prob29}\text{d})}. \notag
\end{align}
where $[\cdot]^+$ denotes $\max\{\cdot,0\}$. \textcolor{black}{
	The penalty factors $\{\upsilon_{1,k}\}$ and $\upsilon_2$ are initialized 
	as $1$ and gradually increased across iterations by multiplying them by a 
	scaling factor $\iota=10$. To determine whether the penalized constraints 
	are sufficiently satisfied, we define the maximum normalized constraint 
	violation as
	\begin{align}\small
	\epsilon_{\mathrm{pen}}
	=
	&\max\Bigg\{
	\max_{k}
	\frac{
		\left[
		\left\|
		\mathbf{F}_{\mathrm{RF},k}
		\mathbf{F}_{\mathrm{D},k}
		\right\|_F^2
		-P_{\mathrm{th}}^{\mathrm{O}}
		\right]^+
	}{
		P_{\mathrm{th}}^{\mathrm{O}}
	},
	\notag\\
	&
	\frac{
		\left[
		\frac{\Gamma}{T}
		-\sum_{k=1}^{K}
		\mathrm{Tr}\left(
		\mathbf{A}_k
		\mathbf{F}_{\mathrm{RF},k}
		\mathbf{F}_{\mathrm{D},k}
		\mathbf{F}_{\mathrm{D},k}^{H}
		\mathbf{F}_{\mathrm{RF},k}^{H}
		\mathbf{A}_k^{H}
		\right)
		\right]^+
	}{
		\Gamma/T
	}
	\Bigg\}.
	\label{penalty_violation}
	\end{align}
	$\quad$The penalty update is terminated when 
	$\epsilon_{\mathrm{pen}}\leq 10^{-4}$; otherwise, the penalty factors are 
	multiplied by $\iota$ and the optimization is repeated. With the iterative 
	increase of the penalty factors, constraint violations are progressively 
	suppressed, driving the obtained solution toward the feasible region of the 
	original problem \cite{xyiot1,xytcom2}.
} Based on the above transformation, the feasible region of subproblem~(\ref{prob20}) becomes a complex circle manifold~$\mathcal{F}=\{\mathbf{F}_{\mathrm{RF},k}\big||[\mathbf{F}_{\mathrm{RF},k}]_{i,j}|=1,\forall i,j\}$. Accordingly, subproblem~(\ref{prob20}) is recast as an unconstrained optimization problem over the manifold $\mathcal{F}$, enabling the application of Riemannian optimization methods. The tangent space of the manifold at the point $\widehat{\mathbf{F}}_{\mathrm{RF},k}$ is given by
\begin{equation}
S_{\widehat{\mathbf{F}}_{\mathrm{RF},k}}\mathcal{F}
=
\left\{
\mathbf{Z} \in \mathbb{C}^{M \times M_{\mathrm{RF}}}
\;\middle|\;
\text{Re}\!\big\{\widehat{\mathbf{F}}_{\mathrm{RF},k} \odot \mathbf{Z}\big\}
=
\mathbf{0}
\right\},
\label{eq:tangent_space}
\end{equation}
where $\mathbf{Z}$ denotes a tangent vector at $\widehat{\mathbf{F}}_{\mathrm{RF},k}$. Denoting the objective function (\ref{prob20}) by $h_{RF}(\mathbf{F}_{\mathrm{RF},k})$, the corresponding Riemannian gradient $\operatorname{grad} h_{RF}$ is obtained by projecting its Euclidean gradient $\nabla h_{RF}$ onto the tangent space $S_{\widehat{\mathbf{F}}_{\mathrm{RF},k}}\mathcal{F}$, i.e.,
\begin{equation}
\operatorname{grad} h_{RF}
=
\nabla h_{RF}
-
\text{Re}\big\{
\nabla h_{RF}\odot \mathbf{F}_{\mathrm{RF},k}^{*}
\big\}
\odot \widehat{\mathbf{F}}_{\mathrm{RF},k}. 
\label{eq:riemannian_grad}
\end{equation}

  \begin{figure*}[!t]	
	\begin{equation} 
	\begin{aligned}\label{prob21}
	\nabla h_{RF} =	& 2\omega_k\mathbf{H}_k^H \mathbf{V}_{\mathrm{RF}}\left(\mathbf{S}\mathbf{V}^H_{\mathrm{RF}}\mathbf{H}_k\widehat{\mathbf{F}}_{\mathrm{RF},k}\mathbf{F}_{\mathrm{D},k}-\mathbf{V}_{\mathrm{D},k}\mathbf{W}_k\right)\mathbf{F}^H_{\mathrm{D},k}+4\upsilon_{1,k}(|\widehat{\mathbf{F}}_{\mathrm{RF},k}\mathbf{F}_{\mathrm{D},k}|^2_F-P^\text{O}_\text{th})\widehat{\mathbf{F}}_{\mathrm{RF},k}\mathbf{F}_{\mathrm{D},k}\mathbf{F}^H_{\mathrm{D},k}\\+&4\upsilon_{2}\left( \sum_{k=1}^{K}\mathrm{Tr}\!\left(\mathbf{A}_k\widehat{\mathbf{F}}_{\mathrm{RF},k}\mathbf{F}_{\mathrm{D},k}\mathbf{F}^H_{\mathrm{D},k}\widehat{\mathbf{F}}_{\mathrm{RF},k}^H\mathbf{A}_k^{H}\right)-\frac{\Gamma}{T}\right) \mathbf{A}_k^{H}\mathbf{A}_k\widehat{\mathbf{F}}_{\mathrm{RF},k}\mathbf{F}_{\mathrm{D},k}\mathbf{F}^H_{\mathrm{D},k}.
	\end{aligned} 
	\end{equation}	
		\hrulefill
\end{figure*}

Defining $\mathbf{S}=\sum_{k=1}^{K}\mathbf{V}_{\mathrm{D},k}\mathbf{W}_k\mathbf{V}_{\mathrm{D},k}^H$, the Euclidean gradient $\nabla h_{RF}$ is given in (\ref{prob21}) at the top of the next page. Then, the variable ${\mathbf{F}}_{\mathrm{RF},k}$ is updated along the Riemannian gradient direction as
\begin{equation} \label{lx1}
\mathbf{F}_{\mathrm{RF},k}^{(n+1)}
=\frac{
	\big(\mathbf{F}_{\mathrm{RF},k}^{(n)} + \psi \operatorname{grad} h_{RF} \big)_{i,j}
}{
	\left|
	\big(\mathbf{F}_{\mathrm{RF},k}^{(n)} + \psi \operatorname{grad} h_{RF} \big)_{i,j}
	\right|
},~\forall i,j,
\end{equation}
where $n$ denotes iteration index and  $\psi$ is the step size.

\subsubsection{Updating $\mathbf{V}_{\mathrm{D},k}$} Given $\{\mathbf{F}_{\mathrm{RF},k},\mathbf{F}_{\mathrm{D},k},\mathbf{V}_{\mathrm{RF}},\mathbf{W}_k\}$,
the optimization problem with respect to the digital combining matrices
$\{\mathbf{V}_{\mathrm{D},k}\}$ is a well-known linear MMSE estimation problem, for which the optimal digital combiner is achieved by the linear MMSE receiver.
Therefore, the optimal $\mathbf{V}_{\mathrm{D},k}$ is given by
\begin{equation} 
\begin{aligned} \label{mmse}
\mathbf{V}_{\mathrm{D},k}^{\star}=\mathbf{J}^{-1}\mathbf{V}_{\mathrm{RF}}^H\mathbf{H}_k\mathbf{F}_{k},
\end{aligned}
\end{equation}
where $\mathbf{J}={\sum_{i=1}^K\mathbf{V}^H_{\mathrm{RF}}\mathbf{H}_{i}\mathbf{F}_{i}(\mathbf{F}_{i})^H\mathbf{H}_{i}^H\mathbf{V}_{\mathrm{RF}}}+{\sigma}^2\mathbf{V}_{\mathrm{RF}}^H\mathbf{V}_{\mathrm{RF}}$ denotes the covariance matrix of the received signal
at the BS. 
\subsubsection{Updating $\mathbf{V}_{\mathrm{RF}}$}The optimization of the analog combining matrix $\mathbf{V}_{\mathrm{RF}}$ proceeds in a manner similar to that of ${\mathbf{F}}_{\mathrm{RF},k}$. Specifically, let $g_{RF}$ denote the objective function of the subproblem (\ref{prob17}) with respect to $\mathbf{V}_{\mathrm{RF}}$. Then, the corresponding Riemannian gradient can be expressed as
\begin{equation}
\operatorname{grad} g_{RF}
=
\nabla g_{RF}
-
\text{Re}\big\{
\nabla g_{RF} \odot \mathbf{V}_{\mathrm{RF}}^{*}
\big\}
\odot \widehat{\mathbf{V}}_{\mathrm{RF}},~\forall i,j. 
\label{eq:riemannian_grad2}
\end{equation}

By defining $\mathbf{C}=\sum_{i=1}^{K}\mathbf{H}_i\mathbf{F}_i\mathbf{F}^H_i\mathbf{H}^H_i+\sigma^2\mathbf{I}$, the Euclidean gradient  $\nabla g_{RF}$ is expressed as
	\begin{equation} 
\begin{aligned}\label{prob22}
 2\sum_{k=1}^{K}\omega_k\mathbf{W}_k\left(\mathbf{C}\widehat{\mathbf{V}}_{\mathrm{RF}}\mathbf{V}_{\mathrm{D},k}\mathbf{V}^H_{\mathrm{D},k}-\mathbf{H}_k{\mathbf{F}}_{\mathrm{RF},k}\mathbf{F}_{\mathrm{D},k}\mathbf{V}^H_{\mathrm{D},k}\right).
\end{aligned} 
\end{equation}

Subsequently, $\mathbf{V}_{\mathrm{RF}}$ is updated along the obtained Riemannian gradient direction as
\begin{equation} \label{lx2}
\mathbf{V}_{\mathrm{RF}}^{(n+1)}
=\frac{
	\big(\mathbf{V}_{\mathrm{RF}}^{(n)} + \psi \operatorname{grad} g_{RF} \big)_{i,j}
}{
	\left|
	\big(\mathbf{V}_{\mathrm{RF}}^{(n)} + \psi \operatorname{grad} g_{RF} \big)_{i,j}
	\right|
}.
\end{equation}

\subsubsection{Updating $\mathbf{W}_k$ } By fixing $\{\mathbf{F}_{\mathrm{D},k},\mathbf{F}_{\mathrm{RF},k},\mathbf{V}_{\mathrm{D},k},\mathbf{V}_{\mathrm{RF}}\}$, problem (\ref{prob17}) reduces to an unconstrained convex problem. The optimal $\mathbf{W}_k$ is obtained by applying the first-order optimality condition, leading to
\begin{equation} 	\label{MMSE3}
\mathbf{W}^*_k=(\mathbf{E}^*_k)^{-1}.
\end{equation}

\textbf{Summary:} By employing the  WMMSE reformulation, the optimization subproblem with respect to $\{\mathbf{F}_k\}_{\forall k}$ and $\{\mathbf{V}_k\}_{\forall k}$ is converted into a WMMSE problem. The variables $\mathbf{F}_{\mathrm{D},k}$, $\mathbf{F}_{\mathrm{RF},k}$,  $\mathbf{V}_{\mathrm{D},k}$, $\mathbf{V}_{\mathrm{RF}}$, and  $\mathbf{W}_k$ are then updated in an alternating manner using SCA, manifold optimization, and closed-form solutions until convergence. The detailed algorithmic procedure for solving subproblem (\ref{prob15}) is summarized in Algorithm \ref{al2}.
\begin{algorithm}[t]
	\renewcommand{\algorithmicrequire}{\textbf{Input:}}
	\renewcommand{\algorithmicensure}{\textbf{Output:}}
	\caption{WMMSE-SCA based optimization algorithm for
		$\{\mathbf{F}_{k}\}_{\forall k}$ and $\{\mathbf{V}_{k}\}_{\forall k}$}
	\begin{algorithmic}[1]
		
		\Require Computation resource allocation
		$\{\boldsymbol{f}^{\mathrm{L}},\boldsymbol{f}^{\mathrm{M}},
		\boldsymbol{f}^{\mathrm{C}}\}$,
		DNN splitting point $l_s$,
		task partition $\{\boldsymbol{\alpha},\boldsymbol{\beta},
		\boldsymbol{\gamma}\}$,
		penalty scaling factor $\iota$,
		penalty violation threshold $\epsilon_{\mathrm{pen}}^{\mathrm{th}}$,
		and convergence threshold $\epsilon$.
		
		\State Initialize $\widehat{R}_k$, $\omega_k$,
		$\upsilon_{1,k}$, and $\upsilon_2$.
		
		\Repeat
		
		\State Update $\mathbf{F}_{\mathrm{D},k}$ by solving problem
		(\ref{prob18}) using the SCA technique.
		
		\Repeat
		\State Update $\mathbf{F}_{\mathrm{RF},k}$ based on (\ref{lx1}).
		\Until{$\left\|\operatorname{grad} h_{\mathrm{RF}}\right\|_F
			\leq \epsilon$}
		
		\State Calculate the maximum normalized constraint violation
		$\epsilon_{\mathrm{pen}}$ according to
		(\ref{penalty_violation}).
		
		\If{$\epsilon_{\mathrm{pen}}
			> \epsilon_{\mathrm{pen}}^{\mathrm{th}}$}
		\State Update
		$\upsilon_{1,k}\leftarrow\iota\upsilon_{1,k},~\forall k$,
		and
		$\upsilon_2\leftarrow\iota\upsilon_2$.
		\EndIf
		
		\State Update $\mathbf{V}_{\mathrm{D},k}$ based on (\ref{mmse}).
		
		\Repeat
		\State Update $\mathbf{V}_{\mathrm{RF}}$ based on (\ref{lx2}).
		\Until{$\left\|\operatorname{grad} g_{\mathrm{RF}}\right\|_F
			\leq \epsilon$}
		
		\State Update $\mathbf{W}_k$ based on (\ref{MMSE3}).
		
		\State Update $\widehat{R}_k=R_k$ and
		$\omega_k=\frac{(\beta_k+\gamma_k)I_k}
		{\widehat{R}_k^2}$.
		
		\Until{the objective value of problem (\ref{prob16}) converges.}
		
		\Ensure Hybrid beamforming and combining matrices
		$\{\mathbf{F}_{k}\}_{\forall k}$ and
		$\{\mathbf{V}_{k}\}_{\forall k}$.
		
	\end{algorithmic}
	\label{al2}
\end{algorithm}

\subsection{Task Partition Optimization}

In this subsection,  we optimize the task partitioning ratios $\bm{\alpha}, \bm{\beta},\bm{\gamma}$ under given $\left\{\bm{f^\text{L}}, \bm{f^\text{M}}, \bm{f^\text{C}}, \{\mathbf{F}_k\}_{\forall k}, \{\mathbf{V}_k\}_{\forall k}\right\}$. The corresponding optimization subproblem can be expressed as
\begin{align}
\hspace*{-3cm} 
\label{prob23}
 \min_{\bm{\alpha}, \bm{\beta},\bm{\gamma}} &  ~~\frac{1}{K}\sum_{k=1}^{K}\max\left\{\,T_k^{\mathrm{L}},~T_k^{\mathrm{E}},~T_k^{\mathrm{C}}\,\right\}\\
\text{s.t.}~&~~ \text{(\ref{prob29}\text{g})},~\text{(\ref{prob29}\text{h})}.\notag
\end{align}

The subproblem (\ref{prob23}) is a linear programming problem and can be directly solved using existing optimization tools such as CVX. To reduce computational complexity, we derive its optimal closed-form solution. Specifically, since the problem  attains its optimal value when $T_k^{\mathrm{L}}=T_k^{\mathrm{E}}=T_k^{\mathrm{C}},\forall k$, the following equations hold:
\begin{equation}
\begin{aligned} \label{prob24} 
\frac{\alpha_kI_kc_k}{f^\text{L}_k}&=\frac{(\beta_k+\gamma_k)I_k}{R_{k}}+\frac{\beta_kI_kc_k}{f^\text{M}_k}, ~\forall k\in \mathcal{K},\\
\frac{\alpha_kI_kc_k}{f^\text{L}_k}&=\frac{(\beta_k+\gamma_k)I_kc_k}{R_{k}}+\frac{\gamma_kI_k}{r_b}+\frac{\gamma_kI_kc_k}{f^\text{C}_k}, ~\forall k\in \mathcal{K}.
\end{aligned}
\end{equation}

By solving (\ref{prob24}) and defining $\widetilde{t}^\text{L}_k=\frac{I_kc_k}{f^\text{L}_k}$, $\widetilde{t}^\text{O}_k=\frac{I_k}{R_{k}}$, $\widetilde{t}^\text{M}_k=\frac{I_kc_k}{f^\text{M}_k}$, and $\widetilde{t}^\text{C}_k=\frac{I_k}{r_b}+\frac{I_kc_k}{f^\text{C}_k}$, we obtain the closed-form solutions of  $\bm{\alpha}$,  $\bm{\beta}$, and $\bm{\gamma}$ as
\begin{equation}
\begin{aligned} \label{prob25}
& \alpha_k^{\star}=\frac{\widetilde{t}^\text{L}_k \widetilde{t}^\text{M}_k}{\widetilde{t}^\text{L}_k \widetilde{t}^\text{M}_k+\widetilde{t}^\text{L}_k \widetilde{t}^\text{C}_k+\widetilde{t}^\text{O}_k \widetilde{t}^\text{M}_k+\widetilde{t}^\text{O}_k \widetilde{t}^\text{C}_k+\widetilde{t}^\text{M}_k \widetilde{t}^\text{C}_k},\\
& \beta_k^{\star}=\frac{\widetilde{t}^\text{L}_k \widetilde{t}^\text{C}_k}{\widetilde{t}^\text{L}_k \widetilde{t}^\text{M}_k+\widetilde{t}^\text{L}_k \widetilde{t}^\text{C}_k+\widetilde{t}^\text{O}_k \widetilde{t}^\text{M}_k+\widetilde{t}^\text{O}_k \widetilde{t}^\text{C}_k+\widetilde{t}^\text{M}_k \widetilde{t}^\text{C}_k}, \\
& \gamma_k^{\star}=\frac{\widetilde{t}^\text{O}_k \widetilde{t}^\text{M}_k+\widetilde{t}^\text{O}_k \widetilde{t}^\text{C}_k+\widetilde{t}^\text{M}_k \widetilde{t}^\text{C}_k}{\widetilde{t}^\text{L}_k \widetilde{t}^\text{M}_k+\widetilde{t}^\text{L}_k \widetilde{t}^\text{C}_k+\widetilde{t}^\text{O}_k \widetilde{t}^\text{M}_k+\widetilde{t}^\text{O}_k \widetilde{t}^\text{C}_k+\widetilde{t}^\text{M}_k \widetilde{t}^\text{C}_k}.
\end{aligned}
\end{equation}
\subsection{Overall Algorithm}
Based on the algorithms introduced in the previous subsections, we now present the overall framework\footnote{\textcolor{black}{Although the proposed AO algorithm requires more than twenty seconds with the CVX toolbox, this runtime mainly comes from the interior-point based solver. In practice, optimization can be accelerated on field programmable gate arrays (FPGA) platforms, achieving microsecond-level execution latency \cite{FPGA}.}} for solving problem (\ref{prob29}). Specifically,  $\{\boldsymbol{f}^\text{L}, \boldsymbol{f}^\text{M}, \boldsymbol{f}^\text{C}\}$ and $l_s$ are  optimized using Algorithm 1,  $\{\mathbf{F}_k\}_{\forall k}$ and  $\{\mathbf{V}_k\}_{\forall k}$ are optimized using Algorithm 2, and  $\bm{\alpha}, \bm{\beta},\bm{\gamma}$ are updated according to (\ref{prob25}). These variables are alternately optimized until convergence. \textcolor{black}{Since problem (\ref{prob29}) is non-convex, the proposed AO algorithm obtains a suboptimal stationary solution rather than the globally optimal solution. In practice, the optimization is performed periodically rather than for each uplink round, and the obtained resource allocation and DNN split decisions can be reused for multiple uplink rounds, thereby amortizing the optimization overhead.}

The overall computational complexity of the proposed framework is mainly dominated by Algorithm 1 and Algorithm 2. The complexity of Algorithm 1 primarily comes from the one-dimensional search and the closed-form updates, which is on the order of $\mathcal{O}(LK)$. The complexity of Algorithm 2 arises from employing the interior-point \cite{2004Convex} method to solve optimization problem (\ref{prob18}), as well as the gradient computation and matrix inversion operations, resulting in a computational complexity of $\mathcal{O}\left(K^3d^3M_{\mathrm{RF}}^3+KMNM_{\mathrm{RF}}+KMNN_{\mathrm{RF}}+N_{\mathrm{RF}}^3\right)$. Therefore, the overall complexity of the proposed algorithm is $\mathcal{O}(it_O(LK+it_IK^3d^3M_{\mathrm{RF}}^3+it_IKMNM_{\mathrm{RF}}+\allowbreak it_IKMNN_{\mathrm{RF}}+it_IN_{\mathrm{RF}}^3))$, where $it_I$ and $it_O$ denote the number of iterations required by Algorithm 2 and the overall algorithm, respectively.

\textcolor{black}{The convergence of the proposed AO algorithm can be guaranteed by the iterative optimization process. Specifically, each subproblem is optimized with the remaining variables fixed, and the obtained solution achieves a non-increasing objective value compared with the previous iteration. Moreover, the objective function is lower bounded by zero due to the non-negativity of latency terms. Therefore, the objective value is monotonically non-increasing over the outer iterations and converges to a finite value.  }

\section{Numerical Results}
In this section, we evaluate the performance of the proposed scheme. We first present the simulation settings, followed by a comprehensive analysis of the simulation results.

{\begin{table}[t]
	\renewcommand{\arraystretch}{1.35}
	\caption{Simulation Parameters.}
	\label{table_example2}
	\centering
	\color{black}{\begin{tabular}{l l}
		\hline
		
		\bfseries Parameters &  \multicolumn{1}{c}{\bfseries Value}\\ 
		\hline
		
		Computational capacity of  cloud server $f^\text{C}_\text{th}$ & 250 Gcycles/s\\
		Computational capacity of the MEC server $f^\text{M}_\text{th}$  & 50 Gcycles/s\\
		Computational capacity of device 
		$k$ $f^\text{L}_\text{th}$ & 5 Gcycles/s\\
		Device $k$'s task  size $I_k$ & 4 MB\\
		Signal bandwidth $B$  & 100 MHz  \\
		Noise power spectral density & -174 dBm/Hz\\
		\tabincell{l}{Maximum transmit power of each device $P^\text{O}_\text{th}$} & 30 dBm\\
		computation intensity $c_k$ & 1000 cycles/bit\\
		FLOPs per CPU cycle at the MEC server $\rho^\text{M}_0$ & 4 FLOPs/cycle \\
		FLOPs per CPU cycle at the cloud server $\rho^\text{C}_0$ & 8 FLOPs/cycle \\
		\tabincell{l}{The transmission rate from MEC server to cloud $r_b$}  & 100 Mbit/s\\
		Initial penalty factors $\{\upsilon_{1,k}\}$, $\upsilon_2$ & 1 \\
		Penalty factor update coefficient $\iota$ & 10 \\ 
		\hline
	\end{tabular}}
\end{table}}
 \subsection{System Setup and Benchmarks}
Unless otherwise specified, the following simulation settings are considered.  The BS is located at $(0\,\text{m}, 0\,\text{m})$ and is equipped with $N=64$ receive antennas and $N_\text{RF}=8$ RF chains. A total of $K=4$ devices are randomly distributed within a rectangular area of $[0\,\text{m}, 100\,\text{m}] \times [-50\,\text{m}, 50\,\text{m}]$, and each device is equipped with $M=8$ transmit antennas and $M_\text{RF}=4$ RF chains. \textcolor{black}{The center of area of interest}  is located at  $[15\,\text{m}, 30\,\text{m}]$.  For the DNN-based sensing task, we consider the classical AlexNet architecture for target classification, whose network structure and parameters are detailed in \cite{Alexnet1}. In addition, we set the  detection probability threshold as  $P_D^{\min} = 0.95$, the false alarm probability as  $P_{\rm {fa}}=1e^{-5}$, and the maximum tolerable latency of the sensing task as $T_\text{th}^\text{I}=8~\text{ms}$. \textcolor{black}{The noise power spectral density is set to -174dBm/Hz. The MEC-cloud transmission rate $r_b$ is set to 100 Mbit/s, which is independent of the uplink transmission rate $R_k$.  The penalty factors $\{\upsilon_{1,k}\}$ and  $\upsilon_2$ in (44) are initialized as 1 and updated by multiplying a factor of $\iota=10$ during iterations until the penalty constraints are sufficiently satisfied. Other key parameters are summarized in Table II.}  To ensure statistical reliability, Monte Carlo simulations are conducted, and the results are averaged over 150 independent runs. To demonstrate the effectiveness of the proposed framework, we  consider the following five benchmark schemes under the same parameter settings for comparison.

\begin{itemize}
	\item {\bfseries Three-tier computing system without integrated sensing (TTWS) \cite{mec4}}: The TTWS scheme focuses on cloud–edge–device collaborative computation and task offloading,  but does not take sensing requirement into account.
	
	\item {\bfseries Fully digital array architecture (FDA) scheme \cite{mec2}}: In the FDA scheme, both the devices and the BS employ fully digital arrays for beamforming and combining.
	
	\item {\bfseries Equal-ratio task partitioning (ERTP) scheme}: In the ERTP scheme, each device task is evenly partitioned among local, MEC server, and cloud computing.
	
	\item {\bfseries Fixed computation resource allocation (FCRA) scheme \cite{peng2024tcom}}: In the FCRA scheme, the computation resources at the MEC and cloud servers are evenly allocated to each device task.
	
	\item {\bfseries Non-split inference (NSI) scheme}: In the NSI scheme, the DNN-based sensing task is wholly executed at the MEC server without model splitting.
\end{itemize}

 \subsection{Convergence Performance}

\begin{figure}[t]
	\centering
	\includegraphics[width=2.7in]{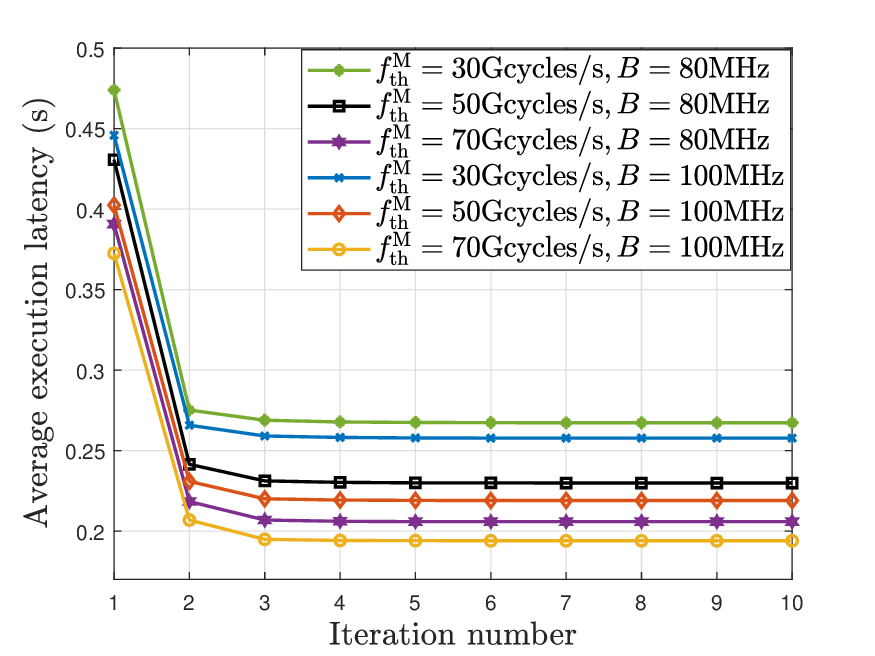}
	\caption{The convergence behavior of overall algorithm.}
	\label{fig:f1}
\end{figure}

In Fig.~\ref{fig:f1}, we present the convergence behavior of the proposed overall algorithm under different MEC server computational capacities $f^{\mathrm{M}}_{\mathrm{th}}$ and signal bandwidths $B$. It can be observed that the algorithm converges within a finite number of iterations for all parameter configurations. Moreover, as $f^{\mathrm{M}}_{\mathrm{th}}$ and $B$ increase, the average task execution latency across all devices decreases accordingly. This is because more available MEC computation resource helps reduce the processing latency of computing tasks, while a larger bandwidth improves the data offloading rate.

 \subsection{The Average Execution Latency and Sensing Performance}

\begin{figure}[t]
	\centering
	\includegraphics[width=2.7in]{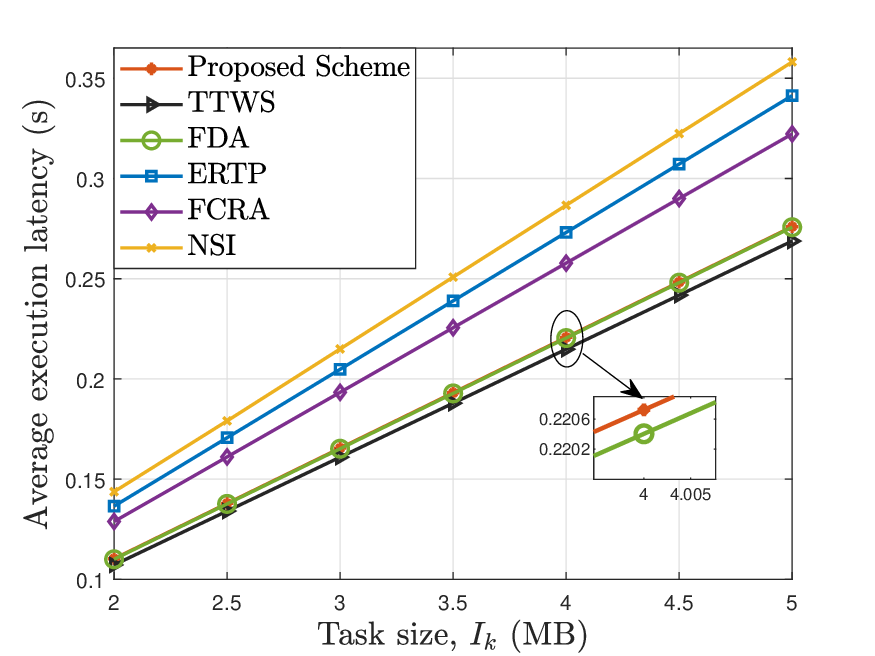}
	\caption{Average execution latency versus task size $I_k$.}
	\label{fig:f2}
\end{figure}

\begin{figure}[t]
	\centering
	\includegraphics[width=2.7in]{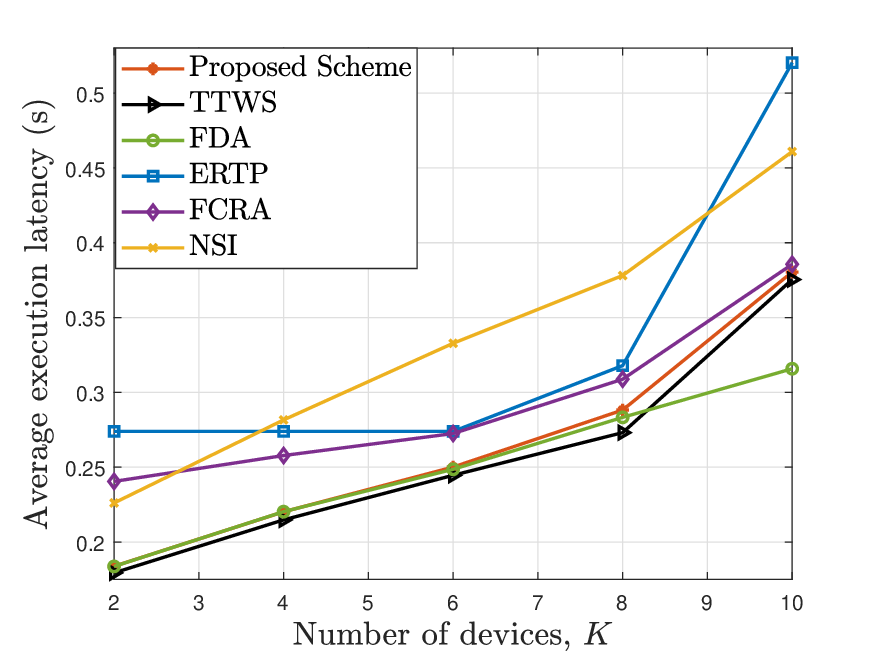}
	\caption{Average execution latency versus number of devices $K$.}
	\label{fig:f3}
\end{figure}

Fig.~\ref{fig:f2} illustrates the average task execution latency across all devices under different schemes. It is observed that the latency increases approximately linearly with the task size $I_k$ for all schemes. Moreover, the proposed framework consistently achieves a lower latency than ERTP, FCRA, and NSI schemes. In ERTP scheme, computation tasks are proportionally distributed among the cloud, MEC server, and devices, making the system more susceptible to bottlenecks caused by specific tiers, such as limited local computing capability or constrained edge–cloud transmission rates. By contrast, the proposed framework adaptively adjusts task partitioning ratios according to the processing capabilities of different tiers, thereby reducing the overall latency. Compared with the FCRA scheme, which allocates identical computation resources to all users, the proposed framework performs computation resource allocation based on heterogeneous channel conditions and device capabilities, leading to higher overall system utility. Furthermore, by incorporating split inference, it enables cooperative task execution between the cloud and MEC servers, alleviating the computation burden at the MEC server. The optimized partition point also reduces the size of forwarded feature data, resulting in a lower latency than the NSI scheme. In addition, the performance of the proposed scheme approaches that of the TTWS and FDA schemes. The comparison with TTWS indicates that joint hybrid beamforming and computation resource allocation can effectively mitigate the latency induced by sensing functionalities, while the comparison with FDA demonstrates the effectiveness of the proposed hybrid beamforming design in approaching fully digital array performance.

\begin{table*}[!t]
	\caption{Computational complexity comparison of different schemes.}
	\label{tab:complexity}
		\renewcommand{\arraystretch}{1.55}
	\centering
	\color{black}{\begin{tabular}{c|c}
		\hline
		\textbf{Scheme} & \textbf{Computational Complexity} \\
		\hline
		TTWS &
		$\mathcal{O}\left(it_{O}(K+it_{I}K^3d^3M_{RF}^3+
		it_{I}KMNN_{RF}+it_IKMNM_{\mathrm{RF}}+it_{I}N_{RF}^3)\right)$
		\\
		\hline
		FDA &
		$\mathcal{O}\left(it_{O}(LK+it_{I}K^3d^3M^3
		+it_{I}N^3)\right)$
		\\
		\hline
		ERTP &
		$\mathcal{O}\left(it_{O}(LK+
		it_{I}K^3d^3M_{RF}^3+
		it_{I}KMNN_{RF}+it_IKMNM_{\mathrm{RF}}+it_{I}N_{RF}^3)\right)$
		\\
		\hline
		FCRA &
		$\mathcal{O}\left(it_{O}(L+
		it_{I}K^3d^3M_{RF}^3+
		it_{I}KMNN_{RF}+it_IKMNM_{\mathrm{RF}}+it_{I}N_{RF}^3)\right)$
		\\
		\hline
		NSI &
		$\mathcal{O}\left(it_{O}(K+
		it_{I}K^3d^3M_{RF}^3+
		it_{I}KMNN_{RF}+it_IKMNM_{\mathrm{RF}}+it_{I}N_{RF}^3)\right)$
		\\
		\hline
		Proposed &
		$\mathcal{O}\left(it_{O}(LK+
		it_{I}K^3d^3M_{RF}^3+
		it_{I}KMNN_{RF}+it_IKMNM_{\mathrm{RF}}+it_{I}N_{RF}^3)\right)$
		\\
		\hline
	\end{tabular}}
\end{table*}

In Fig.~\ref{fig:f3}, we present the average execution latency versus the number of devices $K$. As $K$ increases, the latency of all schemes generally rises, while the proposed algorithm consistently approaches the performance of the TTWS scheme. For the ERTP scheme, which adopts proportional task partitioning, the latency remains nearly unchanged for small $K$, as it is primarily constrained by limited local computing capability. However, when $K\geq8$, the reduced offloading rate becomes the dominant bottleneck, resulting in a noticeable increase in latency. By contrast, the proposed scheme is less sensitive to variations in $K$ owing to its flexible task partitioning capability. Moreover, as $K$ grows, the performance of the FCRA scheme gradually approaches that of the proposed scheme, since the latency becomes dominated by offloading delay and the gain from computation resource allocation diminishes. In addition, the FDA scheme exhibits the smallest latency variation with increasing $K$, owing to its sufficient RF chains that enable beamforming and reception with higher spatial degrees of freedom, thereby mitigating the interference introduced by a larger number of devices.

\textcolor{black}{Furthermore, to evaluate the computational overhead introduced by added sensing mechanisms, we provide a complexity comparison among the proposed scheme and benchmark schemes in Table \ref{tab:complexity}. Although the proposed framework introduces additional optimization components, the complexity increase remains limited by exploiting the closed-form solutions for computation resource allocation and task partition variables, as well as the finite search space of DNN split points. Moreover, by adopting a hybrid beamforming architecture with a limited number of RF chains, the proposed scheme reduces the dimension of beamforming optimization variables and achieves lower computational complexity compared with the FDA scheme.}

\begin{figure}[t]
	\centering
	\includegraphics[width=2.7in]{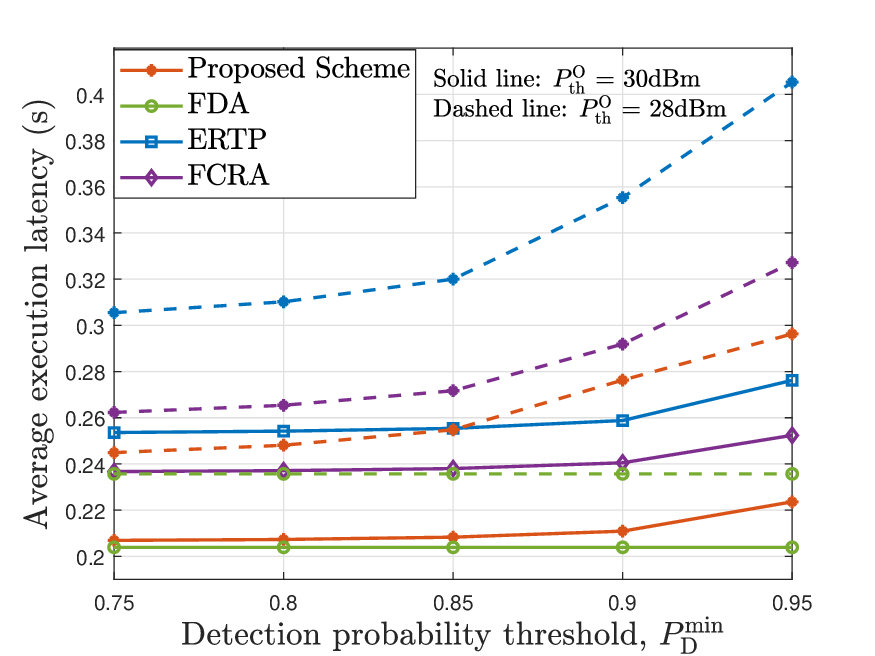}
	\caption{Average execution latency versus detection probability $P^\text{min}_\text{D}$.}
	\label{fig:f4}
\end{figure}

Fig.~\ref{fig:f4} illustrates the average execution latency under different detection probability thresholds $P^{\mathrm{min}}_{\mathrm{D}}$. It is observed that the latency increases with $P^{\mathrm{min}}_{\mathrm{D}}$, since achieving a higher detection probability requires the ISAC beam to place greater emphasis on sensing, which degrades the offloading rate and consequently increases the task execution latency. This reveals an inherent tradeoff between sensing and computation performance. By contrast, the FDA scheme is almost insensitive to variations in $P^{\mathrm{min}}_{\mathrm{D}}$, owing to its larger number of RF chains and higher spatial degrees of freedom, which enable more efficient simultaneous sensing and task offloading. In addition, as the available transmit power $P^{\mathrm{O}}_{\mathrm{th}}$ decreases, the execution latency increases and becomes more sensitive to $P^{\mathrm{min}}_{\mathrm{D}}$. This is because a lower transmit power budget forces the ISAC beam to place greater emphasis on sensing to satisfy the detection requirement, thereby reducing the effective offloading rate.

\begin{figure}[t]
	\centering
	\includegraphics[width=2.7in]{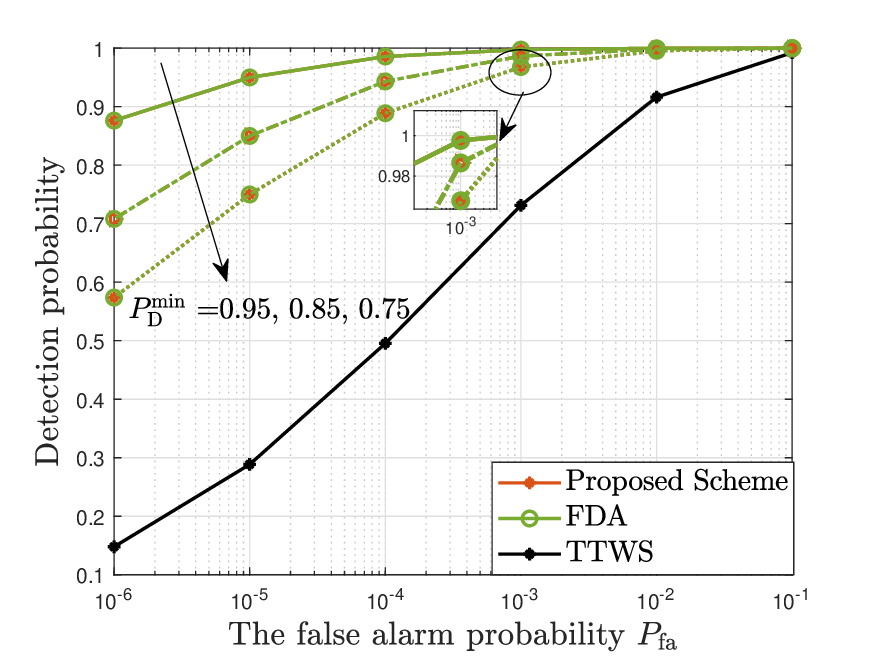}
	\caption{Detection probability versus the false alarm probability  $P_{\rm {fa}}$.}
	\label{fig:f8}
\end{figure}

Fig.~\ref{fig:f8} shows the trade-off between detection probability and false alarm probability of the proposed scheme, FDA, and TTWS.  Benefiting from the explicit consideration of the target detection constraint, the proposed scheme achieves a higher detection probability than TTWS and exhibits almost the same detection performance as FDA. Moreover, as the required detection probability threshold $P_{\rm D}^{\min}$ increases, the achieved detection probability also improves, indicating that the proposed scheme can adaptively adjust the beamforming design to meet more stringent sensing requirements.

\begin{figure}[t]
	\centering
	\includegraphics[width=2.7in]{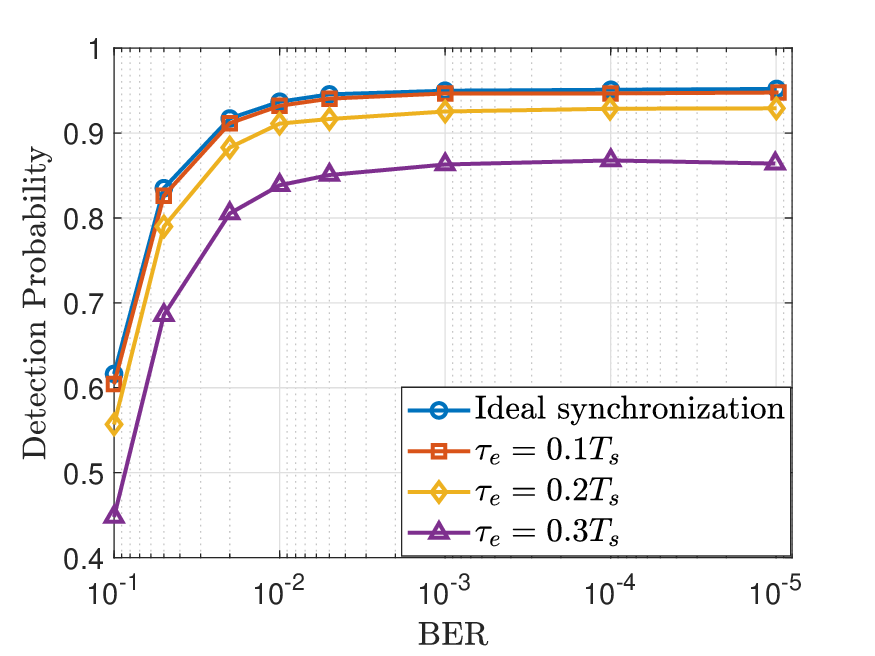}
	\caption{\textcolor{black}{Detection probability versus decoding BER.}}
	\label{fig:f10}
\end{figure}

\textcolor{black}{Fig.~\ref{fig:f10} shows the impact of decoding BER and residual timing errors $\tau_s$ on sensing performance. It can be observed that the detection probability increases as the BER decreases, since more accurate symbol reconstruction provides a more reliable matched-filter reference. When the BER is below $10^{-2}$, the performance degradation caused by decoding errors becomes negligible. Moreover, the impact of residual synchronization errors is evaluated by introducing fractional timing offsets $\tau_s$ relative to the symbol duration $T_s$. We observe that small timing errors only cause limited degradation, while  larger timing offsets lead to more noticeable performance loss due to the mismatch between the received signal and the matched-filter reference.}

\begin{figure}[t]
	\centering
	\includegraphics[width=2.7in]{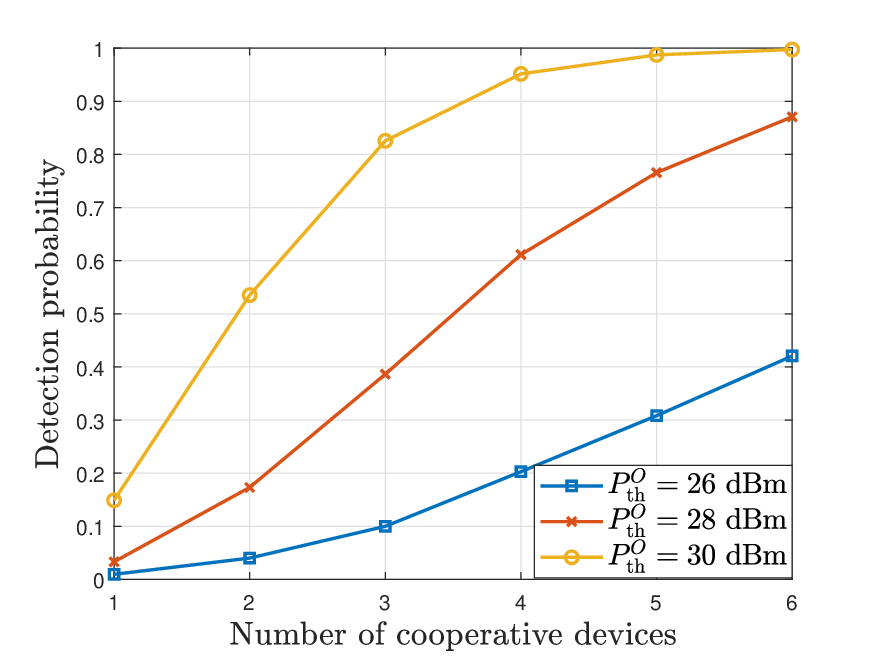}
	\caption{\textcolor{black}{Detection probability versus number of cooperative devices.}}
	\label{fig:f9}
\end{figure}

\textcolor{black}{To further demonstrate the effectiveness of cooperative sensing, we investigate the impact of the number of cooperative devices on the detection performance. As shown in Fig.~\ref{fig:f9}, the detection probability increases significantly with the number of cooperative devices. This is because more cooperative devices provide additional uplink sensing signals, enabling BS to exploit more sensing observations and improve the detection capability. Moreover, a higher transmit power threshold leads to better sensing performance due to the enhanced received signal quality. These results verify the effectiveness of the proposed cooperative sensing mechanism, where multiple devices collaboratively contribute to target detection.}

\begin{figure}
	\setlength{\abovecaptionskip}{-0.1 cm}
	\setlength{\belowcaptionskip}{-0.1cm}
	\centering
	\begin{subfigure}[]
		{\centering
			\includegraphics[width=2.7in]{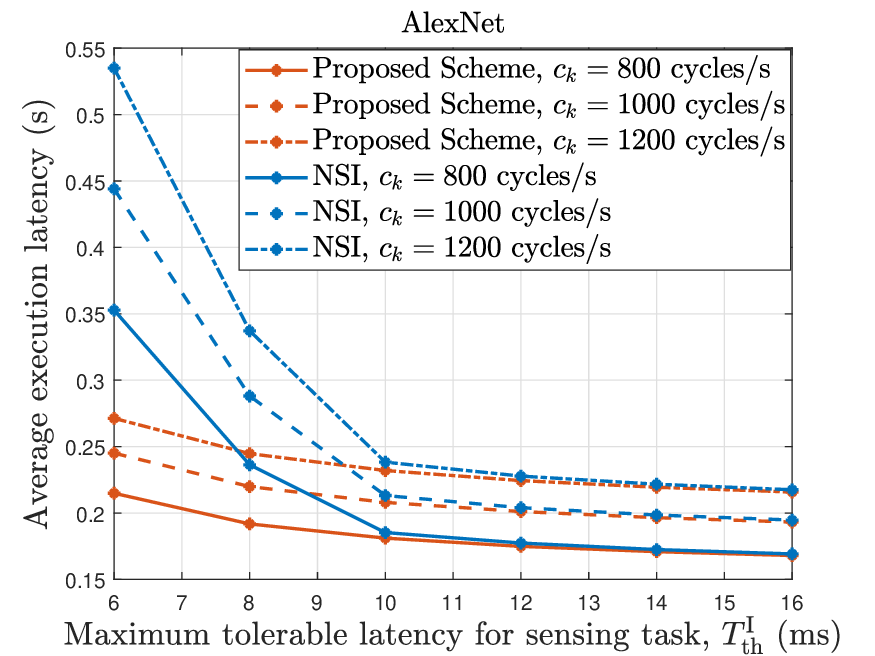}}
	\end{subfigure}
	\hspace{-7 mm}
	\begin{subfigure}[]
		{\centering
			\includegraphics[width=2.7in]{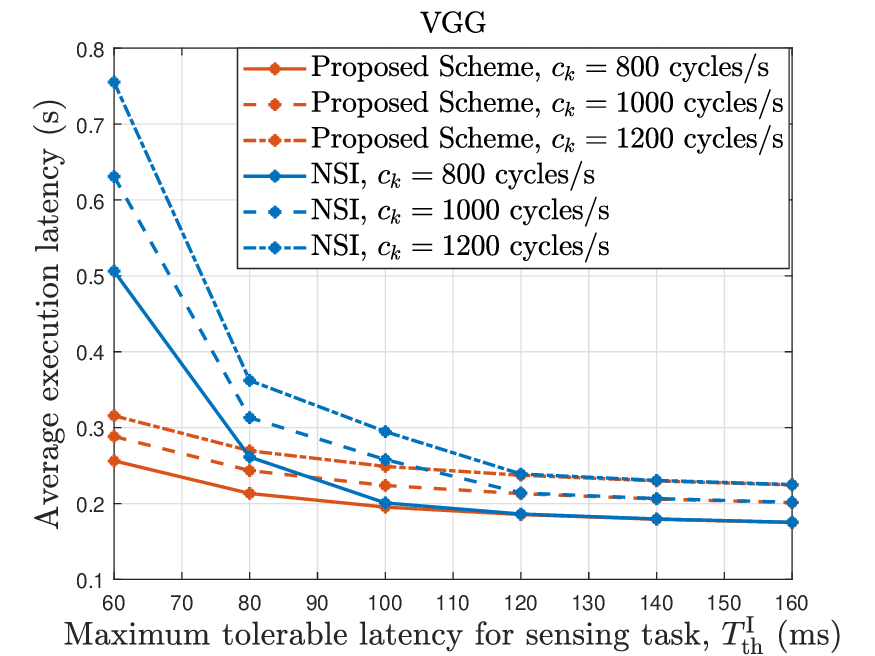}}
	\end{subfigure}
	\caption{\textcolor{black}{Average execution latency versus maximum tolerable latency $T_\text{th}^\text{I}$ of the sensing task. (a) AlexNet. (b) VGG. }}
	\label{fig:f5}
\end{figure}
 
 \begin{table*}[t]
 	\centering
 	\caption*{TABLE IV: Classification accuracy under different DNN split points and intermediate-feature quantization levels.}
 	\label{tab:R4}
 	\renewcommand{\arraystretch}{1.35}
 	\color{black}{\begin{tabular}{c|c|cccc}
 			\hline
 			Split point & Feature dimension & 32-bit & 16-bit & 8-bit & 2-bit \\
 			\hline
 			Non-split baseline & -- & 99.56\% & 99.56\% & 99.56\% & 99.56\%\\
 			feat2 & $27\times27\times96$ & 99.56\% & 99.56\%& 99.33\% & 52.89\% \\
 			feat5 & $13\times13\times384$ & 99.56\% & 99.56\%& 99.56\% & 70.00\% \\
 			feat12 & $6\times6\times256$ & 99.56\% & 99.56\% & 99.56\%& 94.67\% \\
 			avgpool & $6\times6\times256$ & 99.56\% & 99.56\%& 99.56\% & 94.67\% \\
 			fc6 & $4096$ & 99.56\% & 99.56\%& 99.56\% & 96.00\% \\
 			fc7 & $4096$ & 99.56\% & 99.56\%& 99.56\% & 95.50\% \\
 			\hline
 	\end{tabular}}
 \end{table*}
 
To further illustrate the tradeoff between computation and sensing, Fig.~\ref{fig:f5} presents the device task execution latency versus the maximum tolerable sensing latency $T^{\mathrm{I}}_{\mathrm{th}}$ \textcolor{black}{under different DNN architectures including AlexNet and VGG.} It is observed that the execution latency decreases as $T^{\mathrm{I}}_{\mathrm{th}}$ increases, since a relaxed sensing latency constraint allows the MEC and cloud servers to allocate more computation resources to device tasks. Moreover, the proposed scheme is less sensitive to variations in $T^{\mathrm{I}}_{\mathrm{th}}$ than the NSI scheme, owing to its cloud–edge cooperative inference enabled by flexible split design, whereas NSI relies on fully MEC-based processing. When $T^{\mathrm{I}}_{\mathrm{th}}$ is sufficiently large, the performance gap between the two schemes becomes smaller, since only a small portion of MEC computation resources is required for sensing, making split inference less beneficial. \textcolor{black}{Similar performance trends are observed for both AlexNet and VGG architectures, indicating that the effectiveness of the proposed split-inference framework is not limited to a specific DNN architecture.} Furthermore, both the performance gain of the proposed scheme over the NSI scheme and the average execution latency increase with the computation intensity of device tasks $c_k$.

\textcolor{black}{To evaluate the impact of intermediate-feature quantization on the classification accuracy of split inference, we further conduct simulations under different DNN split points and quantization bit-widths. In Table IV. we can observe that, with 32-bit and 16-bit intermediate-feature representations, all split points achieve the same classification accuracy as the non-split baseline (99.56\%). This verifies that the split inference does not affect the classification performance when sufficiently accurate intermediate-feature transmission is adopted. However, when the intermediate features are aggressively quantized to 2 bits, the classification accuracy decreases due to the information loss introduced by quantization. Moreover, earlier split points (e.g., feat2 and feat5) are more sensitive to low-bit quantization because they contain larger and less abstract intermediate representations, while deeper split points exhibit better robustness. These results demonstrate that the accuracy invariance assumption holds under sufficiently accurate intermediate-feature transmission, whereas the impact of feature compression should be considered for practical MEC-cloud deployment.}

%

\section{Conclusion}
This paper investigated a novel ISCC architecture that enables passive sensing by leveraging uplink offloading signals  in the cloud-edge-device collaborative computing system. The results revealed that sensing introduces non-negligible computation latency due to the strong coupling among communication, computation, and sensing. To address this issue, we developed an alternating optimization algorithm to minimize the average computation latency under sensing constraints by jointly optimizing ISAC beamforming, DNN splitting, task partitioning, and computation resource allocation. The proposed framework achieves latency performance close to conventional three-tier computing architectures without sensing, while providing a better tradeoff between computation latency and sensing performance than existing benchmarks. These findings indicate that sensing should be treated as a fundamental component requiring cross-layer joint design in ISCC systems.

\appendix

\bibliographystyle{IEEEtran}
\bibliography{biblp/bibfilelp2}

\end{document}